\documentclass[a4paper,11pt]{article}
\pdfoutput=1 % if your are submitting a pdflatex (i.e. if you have
\usepackage{jcappub} % for details on the use of the package, please
\usepackage[T1]{fontenc} % if needed
\usepackage{aas_macros}
\usepackage{comment}
\usepackage{natbib}
\usepackage{hyperref}
\usepackage{booktabs}
\usepackage{subcaption}
\usepackage[normalem]{ulem}
\usepackage[dvipsnames]{xcolor}
\usepackage{xfrac}
\usepackage{soul}
\title{Constraining the primordial power spectrum using a differentiable likelihood}

\author[a, b]{Subarna Chaki,}
\author[c, a]{Andrina Nicola,}
\author[d]{Alessio Spurio Mancini,}
\author[e]{Davide Piras}
\author[a]{and Robert Reischke}

\affiliation[a]{Argelander-Institut für Astronomie, Universität Bonn,\\Auf dem Hügel 71, D-53121 Bonn, Germany}
\affiliation[b]{Universität Hamburg,  Hamburger Sternwarte\\ Gojenbergsweg 112,  21029 Hamburg, Germany}
\affiliation[c]{Jodrell Bank Centre for Astrophysics, Department of Physics and Astronomy, \\The University of Manchester, Manchester M13 9PL, UK}
\affiliation[d]{Department of Physics, Royal Holloway, University of London,\\Egham Hill, Egham TW20 0EX, UK}
\affiliation[e]{Centre Universitaire d’Informatique, Université de Genève,\\7 route de Drize, 1227 Genève, Switzerland}

\emailAdd{subarna.chaki@uni-hamburg.de}
\emailAdd{anicola@uni-bonn.de}
\emailAdd{Alessio.SpurioMancini@rhul.ac.uk}
\emailAdd{davide.piras@unige.ch}
\emailAdd{rreischke@astro.uni-bonn.de}

\abstract{The simplest inflationary models predict the primordial power spectrum (PPS) of curvature perturbations to be nearly scale-invariant. However, various other models of inflation predict deviations from this behaviour, motivating a data-driven approach to reconstruct the PPS and constrain its shape. In this work, we present a novel method that employs a fully differentiable pipeline to reconstruct the PPS using Gaussian Processes and uses neural network emulators for fast and differentiable theoretical predictions. By leveraging gradient-based sampling techniques, such as Hamiltonian Monte Carlo, our approach efficiently samples the high-dimensional parameter space of cosmological parameters and the free-form PPS, enabling joint constraints on both. Applying this framework to \textit{Planck} 2018 Cosmic Microwave Background (CMB) temperature anisotropy data we find our reconstructed PPS to be consistent with near scale-invariance on small scales, while exhibiting large uncertainties at large scales, driven mostly by cosmic variance. Our results show an overestimation of the PPS amplitude compared to $\Lambda$CDM predictions from the \textit{Planck} 2018 analysis, which we attribute to our choice of a wider prior on the optical depth $\tau$ based on \textit{Planck} 2015 measurements. Adopting a prior consistent with \textit{Planck} 2018 measurements brings our results into full agreement with previous work. To ensure robustness of our results, we validate our differentiable pipeline against a non-differentiable framework, and also demonstrate that our results are insensitive to the choice of Gaussian process hyperparameters. These promising results and the flexibility of our pipeline make it ideally suited for application to additional data sets such as CMB polarisation as well as Large-Scale Structure probes, thus moving towards multi-probe primordial power spectrum reconstruction.}

\begin{document}
\maketitle
\flushbottom
\section{Introduction}\label{sec:intro}
Constraining the physics of the early Universe is one of the key goals in cosmology, and various models have been proposed to describe the processes governing these early times and high energies. Among these, inflation has emerged as the most compelling framework \cite{Guth:1980zm,LINDE1982389,STAROBINSKY198099}, and it now constitutes one of the key pillars of the standard cosmological model, $\Lambda$CDM. According to the theory of inflation, quantum fluctuations during an early period of accelerated expansion give rise to initial perturbations, which evolve into the structures we see in the Universe today. The statistical properties of these fluctuations are encoded in the primordial power spectrum (PPS), making it a vital probe for understanding inflationary physics. These early perturbations are also imprinted on cosmological observables, with the Cosmic Microwave Background (CMB) serving as a particularly powerful tool for studying the PPS. This is largely due to the CMB’s high-precision measurements and its ability to probe a wide range of spatial scales, providing strong constraints on the physics of the early Universe~\cite{planckoverview2018,2019arXiv190712875P}

The simplest models of inflation, particularly canonical single-field slow-roll models \cite{LINDE1982389}, predict initial scalar\footnote{While inflation also predicts the generation of tensor perturbations, we only focus on scalar perturbations for the remainder of this work.} perturbations to be adiabatic and Gaussian. These fluctuations are fully described by a nearly scale-invariant PPS with amplitude $A_{s}$ and scale-dependence characterised by the scalar spectral index $n_{s}$. However, more complex models of inflation predict deviations from near scale-invariance and can also lead to distinctive features in the PPS. As a specific example, models with step-like features or periodic oscillations in the inflaton potential can generate corresponding localised features in the PPS \cite{1992JETPL..55..489S,PhysRevD.64.123514,2008PhRvD..77b3514J,2009JCAP...06..028J}. While recent observations have shown remarkable consistency with near-scale-invariance of the PPS (see e.g. Refs.~\cite{2020Planck2018Cosmoparams, 2020A&A...641A..10P}), the physics of inflation remains unknown. Constraining the shape of the PPS thus offers a unique possibility to constrain the dynamics of the early Universe and test different models of inflation. 

Deviations from near-scale invariance can also be constrained by going beyond the standard $A_s$ and $n_s$ parametrization of the PPS. The simplest model extensions allow for a scale-dependence of the scalar spectral index $n_s$, characterised by the so-called running of the spectral index, $\sfrac{\mathrm{d}n_s}{\mathrm{d}\log k}$, or the running of the running, $\sfrac{\mathrm{d}^2n_s}{\mathrm{d}\log k^2}$. Several analyses, such as Refs.~\cite{2003MNRAS.342L..72B,2006JCAP...03..001B,PhysRevD.52.R1739,2016CQGra..33k5008Z,2006JCAP...09..010E,2020A&A...641A..10P}, have found these running terms to be small, consistent with single-field slow-roll inflation. In addition, various studies have explored oscillatory features that might be superimposed on the standard nearly scale-invariant spectrum, which could arise from more complex inflationary models or early-Universe physics~\cite{2012MNRAS.421..369M,2014PhRvD..89f3536M,Meerburg:2013dla}.

Going beyond standard parametric models, the PPS can also be directly reconstructed from observational data. This approach allows for arbitrary shapes of the PPS, free from any assumptions about inflation, and thus facilitates the comparison with a variety of inflationary models. 
A number of different methods for PPS reconstruction have been considered in the literature. One common approach is wavenumber binning, where the PPS is divided into discrete intervals of $k$, with the amplitude estimated independently in each bin \cite{2002ApJ...573....1W,2003MNRAS.342L..72B,2004JCAP...04..002H,2011arXiv1105.4887H}. Alternatively, wavelet expansions decompose the PPS into localized basis functions at different scales \cite{2003ApJ...598..779M,2003ApJ...599....1M}. Smoothing splines provide a flexible, regularized method to fit the PPS with smooth functions \cite{2005PhRvD..72j3520S,2008JCAP...07..009V,2016JCAP...08..028R}. Other techniques include using inversion methods to directly solve the integral equations linking the PPS to cosmological observables (e.g., CMB anisotropies) using transfer functions~\cite{2002PhRvD..65h3007M,2004ApJ...607...32K,PRISM}, Bayesian approaches leveraging probabilistic frameworks for PPS inference~\cite{2012JCAP...06..006V,2012JCAP...10..050G,2019PhRvD.100j3511H,2024JCAP...06..072M}, and deconvolution algorithms such as the Richardson-Lucy method iteratively correcting for the blurring effect of transfer functions, similar to image deblurring techniques~\cite{2004PhRvD..70d3523S,2009JCAP...07..011N,2021JCAP...10..081C,2024JCAP...03..056S}. 

Building upon these analyses, in this work we use Gaussian processes (GPs) to reconstruct the PPS. While not entirely non-parametric, GPs allow for modelling functions with minimal assumptions about their functional form \cite{articleGP} as well as robust characterization of statistical uncertainties. GPs have thus been employed for model-independent reconstruction of cosmological quantities in several works. For instance, they have been used to reconstruct the inflaton’s speed of sound from \textit{Planck} 2018 data~\cite{2021PhRvD.103l3531C}, to perform model-independent cosmography by mapping the cosmic expansion history~\cite{PhysRevD.85.123530}, and to reconstruct the growth factor~\cite{2024limberjack}.

One key challenge in PPS reconstruction is the large number of parameters required to model cosmological observables, which makes traditional Markov chain Monte Carlo (MCMC) algorithms such as Metropolis-Hastings, extremely inefficient. In this work, we address this challenge by developing a fully differentiable pipeline built on neural network (NN) emulators for efficient theoretical predictions. This allows us to employ gradient-based sampling techniques such as Hamiltonian Monte Carlo (HMC) \cite{DUANE1987216,2011hmcm.book..113N} for efficiently constraining the large parameter space spanned by cosmological parameters and the free-form PPS. In this proof-of-concept study, we apply our pipeline to reconstruct the PPS from the CMB temperature power spectrum measured by the \textit{Planck} Collaboration in 2018 \cite{2019arXiv190712875P,2020Planck2018Cosmoparams}, and use it to derive joint constraints on cosmological parameters and the Gaussian process describing the PPS. 

We note that several other differentiable CMB likelihoods and tools are also available, including \texttt{candl}~\cite{2024A&A...686A..10B} and \texttt{clipy}\footnote{\url{https://github.com/benabed/clipy}}, the latter of which is designed to be compatible with \texttt{candl}. These tools offer access to a broader range of datasets, explicit nuisance parameter modelling, and also include CMB lensing, which is particularly valuable for comprehensive cosmological inference. In future work, integrating such tools into our framework would be a natural and valuable extension, enabling analyses that do not rely on pre-marginalized likelihoods and allowing for joint inference of cosmological and nuisance parameters. Since a detailed comparison is beyond the scope of this study, we leave such integration and benchmarking to future work.

This paper is structured as follows: in Section \ref{sec:methods}, we outline our methodology for reconstructing the PPS from CMB temperature anisotropy data using GPs and NN emulators. Section \ref{sec:data_likelihood} is dedicated to describing the data utilised in our study, along with the employed likelihoods. In Section \ref{sec:results}, we present and discuss our results, and conclude in Section \ref{sec:conclusions}, where we also outline potential avenues for future research. Additional details regarding the analysis and methods are deferred to the appendices. 

\section{Methods} \label{sec:methods}
\begin{figure*}[tbp]
\centering % \begin{center}/\end{center} takes some additional vertical space
\includegraphics[width=1.0\textwidth]{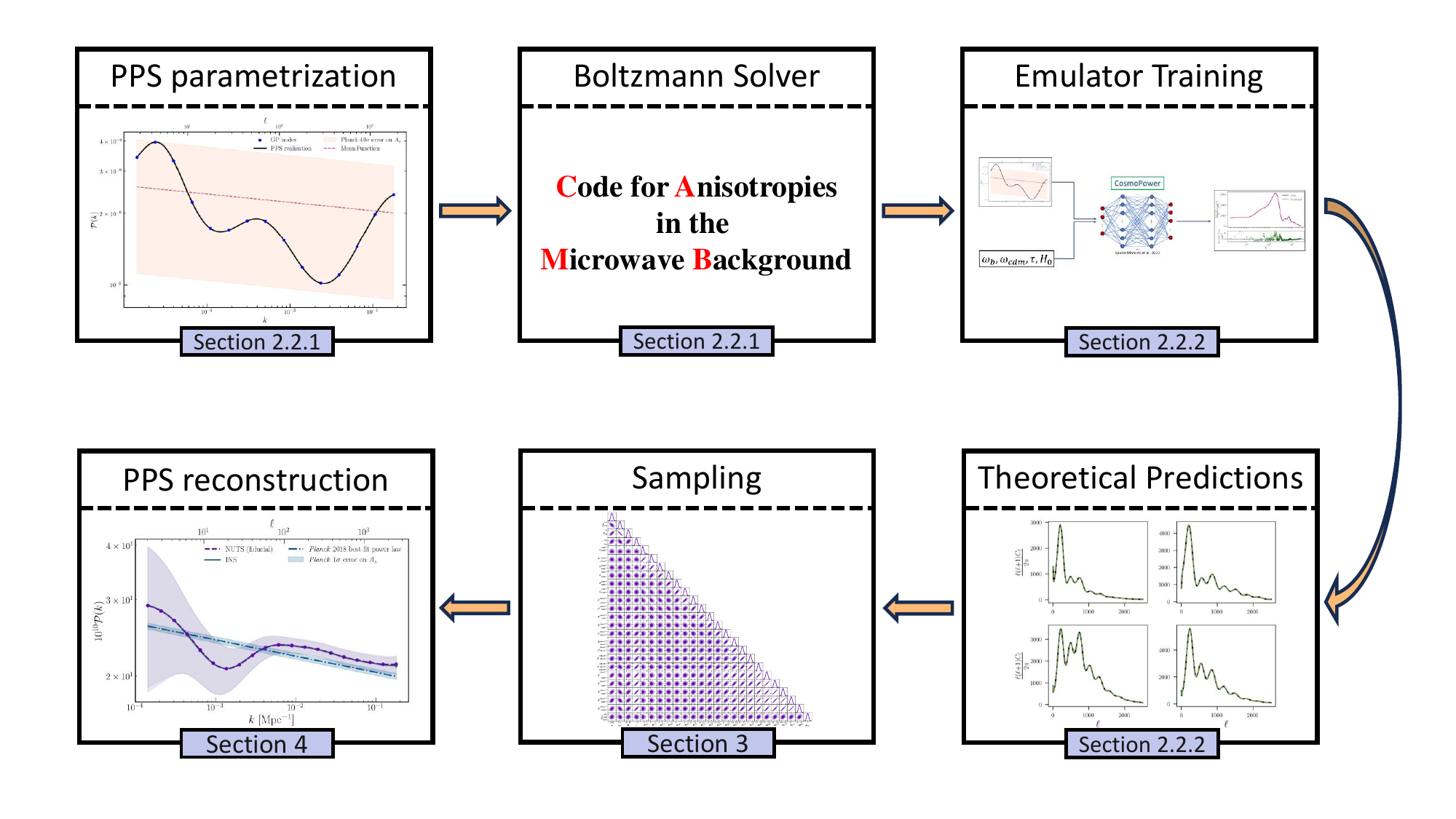}
\caption{\label{fig:flowchart} Overview of the differentiable pipeline for reconstructing the PPS. The top panel illustrates the data generation process and emulator training using \textsc{CosmoPower} (CP). The bottom panel shows the inference pipeline, leading to joint constraints on cosmological parameters and the free-form PPS, thus enabling its reconstruction.}
\end{figure*}

In the following, we describe our analysis pipeline to reconstruct the PPS. The main steps and their dedicated sections are illustrated in Fig.~\ref{fig:flowchart}. 

\subsection{\label{subsec:p(k)}Primordial power spectrum reconstruction}
The primordial scalar power spectrum is most commonly defined in terms of the variance of super-horizon comoving curvature perturbations, $\chi$, in early radiation-domination through 
\begin{equation}
\langle |\chi|^2\rangle=\int\mathrm{d}\log k\;\mathcal{P}_{\chi\chi}(k)\;,
\end{equation}
where $\mathcal{P}_{\chi\chi}(k)$ denotes the dimensionless primordial curvature power spectrum, and $\log$ refers to natural logarithm throughout this work. As discussed in Sec.~\ref{sec:intro}, the simplest models of inflation predict a nearly scale-invariant PPS, which can be parametrized as
\begin{equation}
    \mathcal{P}_{\chi\chi}(k) = A_s\left( \frac{k}{k_0} \right)^{n_s - 1}\;. \label{eq:powerlaw}
\end{equation}
Here, $A_s$ denotes the power spectrum amplitude and $n_s$ is the scalar spectral index, with $n_s=1$ for a scale-invariant spectrum. The quantity $k_0$ denotes a pivot scale, which, following Ref.~\cite{2020Planck2018Cosmoparams}, we set to 0.05 $\text{Mpc}^{-1}$ throughout our analysis. Finally, to ease notation, we will omit the subscript $\chi$ for the remainder of this work, i.e. we set $\mathcal{P}(k)\equiv\mathcal{P}_{\chi\chi}(k)$.

The statistical properties of the primordial fluctuations are reflected in the temperature and polarisation anisotropies of the CMB. While CMB temperature anisotropies are characterised by the scalar field $T$, the spin-2 CMB polarisation field is fully described by the curl-free $E$- and divergence-free $B$-modes. The relation between the PPS, $\mathcal{P}(k)$, and the CMB anisotropy power spectrum, $C^{\mathsf{XY}}_{\ell}$, is established through the transfer function $T^{\mathsf{X}}_{\ell}(k)$, which describes how primordial fluctuations evolve into observed anisotropies. This can be expressed through (see e.g. Ref.~\cite{Dodelson:2003ft})

\begin{equation}
C_{\ell}^{\mathsf{XY}} \propto \int_{}^{}\text{d} \log {{k}}\;\mathcal{P}({k})T^{\mathsf{X}}_{\ell}({k})T^{\mathsf{Y}}_{\ell}({k})\;,
\label{eq:cmb_cl}
\end{equation}
where $\mathsf{X}$ and $\mathsf{Y}$ denote $T$, $E$, or $B$. The transfer function for a given wavenumber $k$ approximately peaks at an angular multipole $\ell$ given by 
\begin{equation}
\ell \simeq k d \;\label{eq:k_l_rel},
\end{equation}
where $d$ denotes the comoving distance to the last scattering surface, which is roughly given by $d = 14,000$ \text{Mpc} for the best-fit cosmological parameters from \textit{Planck} 2018 \cite{2020Planck2018Cosmoparams}.\footnote{We note that, based on the \textit{Planck} 2018 best-fit parameters from Table 2, column 2 (\texttt{TT+lowE}) of Ref.~\cite{2020Planck2018Cosmoparams}, we obtain \( d \simeq 13,876 \)~\text{Mpc}. However, for simplicity and consistency in our analysis, we use $d = 14,000$ \text{Mpc}.}

\begin{figure*}[tbp]
\centering % \begin{center}/\end{center} takes some additional vertical space
\includegraphics[width=0.80\textwidth]{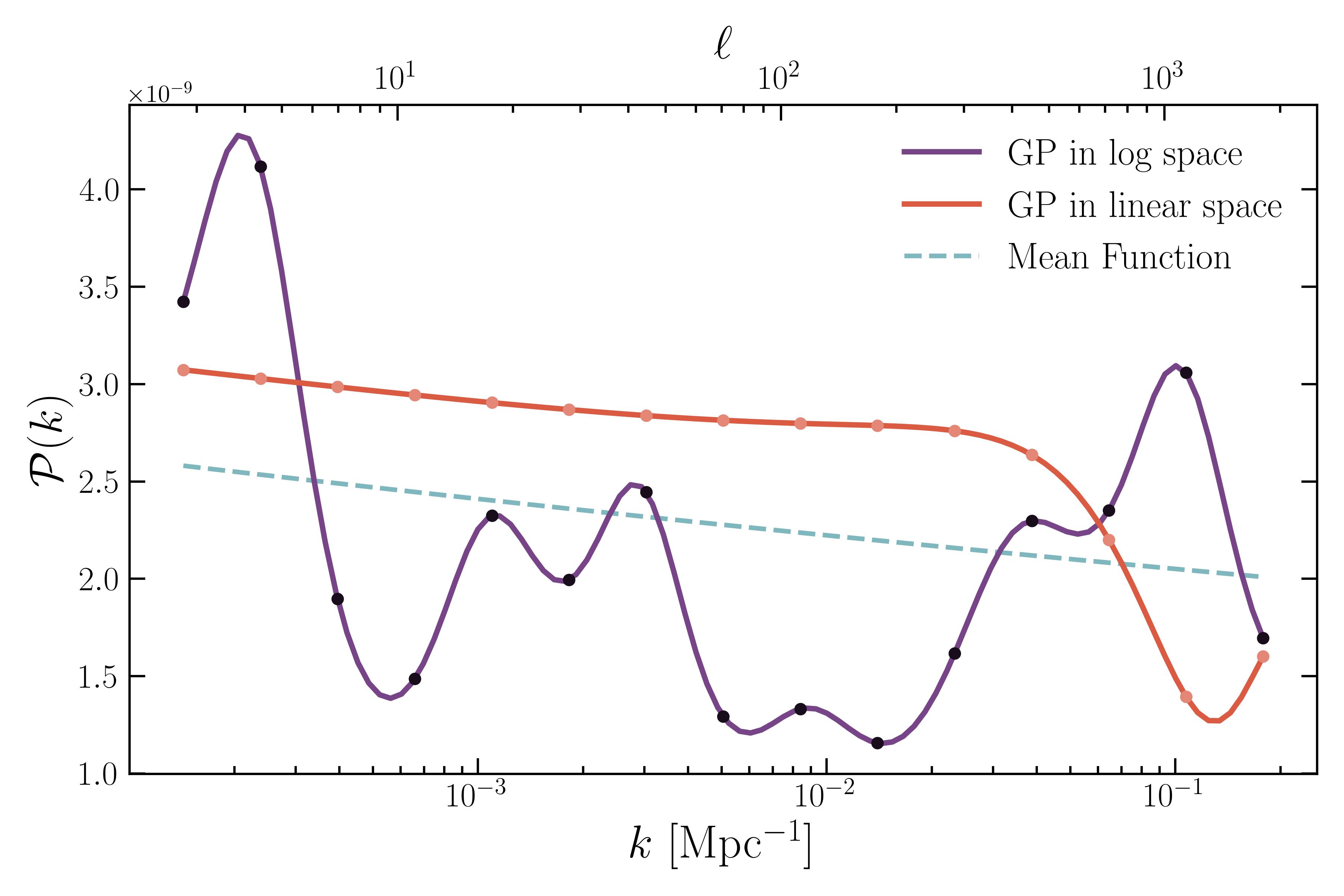}
\caption{\label{fig:linear_vs_log} Comparison of two GP realizations of the PPS (see Eq.~\ref{eq:gp0.1}), generated by imposing correlations between the nodes in linear- and logarithmic-space. Note that the GP hyperparameters used to create the realizations are chosen solely to highlight the differences between the two spaces for clearer visualization; these are not the hyperparameters used in the analysis.}
\end{figure*}

In this work, we employ Gaussian processes to reconstruct the PPS from the CMB temperature power spectrum in order to study deviations from near scale-invariance. 
The main reasons for this choice are twofold. First, GPs allow for reconstructing functions with minimal assumptions on their form except for smoothness, which is a general expectation for the PPS (see e.g. Ref.~\cite{2008JCAP...07..009V,2010PhRvD..81b1302P,2011arXiv1105.4887H}). And second, GPs allow for straightforward quantification of uncertainties, which facilitates the interpretation of the reconstructed PPS (see Section~\ref{sec:results} for more details).

Generally, a GP can be thought of as a framework for describing a distribution of functions, where the values of the function at different points (nodes) are treated as an ensemble of random variables. These are drawn from a multivariate Gaussian distribution governed by a mean function and a covariance function or kernel. The mean function represents the expected value of the process at each node, while the covariance function describes how the values at different nodes are correlated, effectively encoding the smoothness and structure of the process across scales. Therefore, a generic GP at nodes $\mathbf{x}$, $g(\mathbf{x})$, is fully specified by a mean function $m(\mathbf{x})=\langle g(\mathbf{x})\rangle$ as well as a covariance function $K(\mathbf{x}, \mathbf{x}')=\langle(g(\mathbf{x})-m(\mathbf{x}))(g(\mathbf{x}')-m(\mathbf{x}'))^T\rangle$ where $\langle \cdots\rangle$ denotes expectation values. In the following, we will briefly discuss the aspects of GPs most relevant to this work but refer the reader to Ref.~\cite{articleGP} for a comprehensive review. 

In this work, we aim to reconstruct the PPS, $\mathcal{P}(k)$, by modelling the logarithm of the PPS as a GP over a set of nodes in $\log k$-space, i.e., $\log \mathcal{P}(\log k)$. This choice ensures that correlations between nodes are applied in logarithmic space, leading to a constant correlation length (as defined by the covariance kernel in Eq.~\ref{eq:gp2}) across the entire wavenumber range. In contrast, given the range of scales relevant for PPS reconstruction, modelling correlations in linear space causes strong correlations at large scales (small $k$) and weak or negligible correlations at small scales (large $k$). This effect is illustrated in Fig.~\ref{fig:linear_vs_log}, where two GP realizations of the PPS show that correlations between nodes remain constant when imposed in logarithmic space, but decay as a function of $k$ when imposed in linear space. Additionally, modelling $\log \mathcal{P}(\log k)$ naturally prevents unphysical negative values of the PPS, ensuring physical consistency in the reconstruction.

As our mean function, we choose the best-fit power law (see Eq.~\ref{eq:powerlaw}) obtained from \textit{Planck} 2018 \cite{2020Planck2018Cosmoparams}, i.e.
\begin{equation}
\log \mathcal{P}_{\text{P18}}(\log k) = \log(A_s) + (n_s - 1) \left( \log k - \log k_0 \right)\;.
\label{eq:mean_func} 
\end{equation}
We set $A_s = 2.092 \times 10^{-9}$ and $n_s = 0.9626$ based on the results reported by the \textit{Planck} Collaboration in Ref.~\cite{2020Planck2018Cosmoparams}\footnote{Specifically, we take these values from the \texttt{TT+lowE} constraints listed in Table 2, column 2 of Ref.~\cite{2020Planck2018Cosmoparams}.}.
Following the approach of Ref.~\cite{2021PhRvD.103l3531C}, we use a squared exponential kernel as our covariance function, i.e. we set
\begin{equation}
\mathrm{\mathbf{K}}(\log k_i, \log k_{i+1}) = \sigma_f^2 \exp\left(-\frac{(\log k_i - \log k_{i+1})^2}{2 l^2}\right)\;.
\label{eq:gp2}
\end{equation}
The squared exponential kernel depends on two so-called hyperparameters, $l$ and $\sigma_{f}^2$, which control the behaviour of the GP. The correlation length $l$ determines the rate at which the correlation between the function values decays as the distance between the nodes increases, controlling the smoothness of the generated values. The variance $\sigma_f^2$ controls the overall scale of the function values around the mean function and thus effectively sets the spread of the generated realizations\footnote{Throughout this work, we denote the second hyperparameter as $\sigma_f$, representing the standard deviation of the signal, which is the square root of the variance $\sigma_f^2$.}. The choice of using this kernel is motivated by the fact that we expect the PPS to be a continuous function with smooth deviations from its power law form (Eq.~\ref{eq:mean_func}), two properties that are naturally imposed on the GP by our use of the infinitely differentiable exponential kernel.

In our fiducial analysis, we evaluate the above GP at 20 nodes in $k$, logarithmically spaced between $1.429 \times 10^{-4}\ \text{Mpc}^{-1}$ and $1.791 \times 10^{-1}\ \text{Mpc}^{-1}$. This range is calculated using Eq.~\ref{eq:k_l_rel} for the multipole range $\ell = 2$ to $\ell = 2508$ covered by the \textit{Planck} 2018 data \cite{2020Planck2018Cosmoparams,2019arXiv190712875P}. In other words, we choose to reconstruct the PPS only within the scales covered by \textit{Planck} and do not perform any extrapolations. The choice of the number of nodes is arbitrary a priori, and we have investigated the impact of varying this quantity in Appendix~\ref{appendix:n(p(k))}. 

Putting everything together, our full GP PPS is given by
\begin{equation}
    \log \mathcal{P}(\log k) = \log \mathcal{P}_{\text{P18}}(\log k) + \mathbf{L}_{C}(\mathrm{\mathbf{K}}) \cdot \mathbf{z},\label{eq:gp0.1}
\end{equation}
where $ \log \mathcal{P}_{\text{P18}}(\log k)$ is the mean function of our GP given in Eq.~\ref{eq:mean_func}, and $\mathbf{L}_{C}(\mathrm{\mathbf{K}})$ denotes the lower triangular matrix obtained from the Cholesky decomposition of the covariance matrix $\mathrm{\mathbf{K}}$ defined in Eq.~\ref{eq:gp2}. Finally, $\mathbf{z}$ is a vector of random variables, sampled from a standard normal, which we use to generate realizations of the PPS around the mean function. Eq.~\ref{eq:gp0.1} can be interpreted as follows: the matrix $\mathbf{L}_{C}(\mathrm{\mathbf{K}})$ transforms the initially uncorrelated elements of $\mathbf{z}$ into a set of correlated variations that follow the structure specified by the covariance function $\mathbf{K}$. By combining these correlated variations with the mean function $ \log \mathcal{P}_{\text{P18}}(\log k)$, we obtain GP realizations of the PPS that smoothly vary across scales. Thus, in this model, reconstructing the PPS is equivalent to constraining the values of the random vector $\mathbf{z}$. 

As has been noted in several works (see e.g. Refs.~\cite{2022MNRAS.512.1967R,2022PhRvD.106h3523R,2023JCAP...02..014H}), the choice of the GP mean can bias the final reconstruction. This is of particular concern in regions where there is little data, in which the GP is naturally drawn towards its mean function. We do not expect this to be a major issue in our analysis as, given previous measurements, we anticipate only small deviations of the reconstructed PPS from the assumed power law (Eq.~\ref{eq:mean_func}). In addition, we manifestly restrict our reconstruction to regions covered by the observational data. Nevertheless, in Sec.~\ref{sec:results}, we test the robustness of our results to changes in the mean function.

\subsection{\label{predict_cmb}Theoretical predictions}
As seen in Eq.~\ref{eq:cmb_cl} the CMB power spectrum depends both on the PPS and the cosmological parameters that determine the shape of the transfer function, leading to potential degeneracies and thus necessitating a joint estimation. Jointly estimating these cosmological parameters and the nodes of the PPS GP (in our fiducial setup we use 20 nodes), presents a significant computational challenge due to the high dimensionality of the parameter space. 

To address this challenge, we use emulators to compute theoretical predictions. Essentially, emulators are NNs that approximate the output of computationally-intensive functions, such as Boltzmann solvers in our case. These types of emulators have become essential in cosmology in order to sample the large parameter space needed to describe observables in a reasonable time (see e.g. Refs.~\cite{Jimenez:2004ct, 2008MNRAS.387.1575A, 2014ApJ...780..111H, 2020MNRAS.497.2213M,2020MNRAS.491.2655M,2021arXiv210414568A,2022CP,CPJ2023OJAp....6E..20P,2024OJAp....7E..10B,2023JCAP...05..025N,2025arXiv250313183G}). In this work, we use the popular CP framework \cite{2022CP} for building our emulators. An additional advantage of emulators is that, being NNs, they are differentiable by definition, and thus allow for the use of gradient-based sampling methods, enabling us to explore the large parameter space more effectively.  

\subsubsection{Emulator} \label{sec:data_gen} 
\begin{table}[tbp]
\centering
\renewcommand{\arraystretch}{1.1} 
\setlength{\tabcolsep}{12pt}
\begin{tabular}{c|c}
\toprule
\textbf{Parameter} & \textbf{Prior Distribution} \\
\midrule
$\omega_b$         & $\mathcal{U}[0.015,\, 0.03]$    \\
$\omega_{cdm}$     & $\mathcal{U}[0.05,\, 0.3]$      \\
$H_0$              & $\mathcal{U}[50,\, 90]$         \\
$\tau$             & $\mathcal{U}[0.01,\, 0.3]$      \\
GP nodes           & $\mathcal{N}(0,\, 1)$           \\
\bottomrule
\end{tabular}
\caption{\label{tab:prior_tab} Prior distributions for the cosmological parameters and GP nodes used in the NN emulators described in Section~\ref{sec:data_gen}. Here, $\mathcal{U}[a, b]$ denotes a uniform distribution and $\mathcal{N}(0,\, 1)$ a standard normal distribution.}
\end{table}
In this work, we model the CMB temperature power spectrum as a function of four cosmological parameters: the physical baryon density, $\omega_b$, the physical cold dark matter density, $\omega_\mathrm{cdm}$, the Hubble parameter today, $H_0$, and the optical depth to reionization, $\tau$. In our fiducial setup, we additionally allow the PPS to vary at the 20 discrete $k$-nodes as described in Section~\ref{subsec:p(k)}, and denote these by $\left\{ \mathcal{P}_i \right\}_{i=1}^{20}$, where $\mathcal{P}_i$ represents the power spectrum at the $i$-th $k$-mode\footnote{Although the sampling is performed in terms of $\mathbf{z}$ (see Eq.~\ref{eq:gp0.1}), the results presented in Section~\ref{sec:results} are expressed in terms of $\mathcal{P}_i$. This choice facilitates a direct visual comparison between the reconstructed primordial power spectrum and the corresponding contours shown in the corner plots.}. The remaining parameters specifying our cosmological model are kept constant throughout our analysis: in particular we assume a flat universe (i.e. we set the curvature density parameter $\Omega_{K}$ to 0), and assume dark energy to be described by a cosmological constant. Furthermore, we assume one massive neutrino with mass of 0.06 $\text{eV}$. 

We train a NN emulator to learn the mapping between these 24 input parameters and the CMB temperature power spectrum at angular multipoles \(\ell\) ranging from 2 to 2508 as computed using the Boltzmann code \textsc{camb} \cite{2000CAMB}. We use standard \textsc{camb} accuracy settings, which have been verified for \textit{Planck} precision, and as discussed in Appendix~\ref{appendix:nn_train} find these to be sufficient for the purpose of our analysis. \textsc{camb} uses the cosmological parameters and $\mathcal{P}_i$, as inputs to compute the corresponding CMB temperature power spectrum. To input our PPS into \textsc{camb}, we first subsample the 20 parametrized PPS values onto a finer grid of 100 nodes using our GP model, and then create a third-order spline interpolator, which we finally pass to the Boltzmann solver.

To ensure the robustness of this approach, we have tested various interpolation methods (including performing the interpolation within \textsc{camb}) and found that the choice of method does not significantly affect the results in our analysis.

To generate the training data for the NN emulator, we follow the workflow described in Ref.~\cite{2022CP}\footnote{For a comprehensive tutorial, please refer to the GitHub repository: \href{https://github.com/alessiospuriomancini/cosmopower.git}{\textsc{CosmoPower}}.}. We sample the cosmological parameters and PPS nodes using Latin hypercube sampling (LHS) within the prior ranges specified in Tab.~\ref{tab:prior_tab}. We note that as the priors on the PPS nodes are Gaussian, we use inverse transform sampling to map the LH samples to standard normal random variables. Our training set consists of approximately \(6 \times 10^4\) samples\footnote{Note that it took $\sim 2$ hours to generate the samples on 48 CPU cores. Also, the exact number of samples may vary, as certain parameter combinations from the hypercube can lead to integration timeouts during the computation of the CMB temperature power spectrum using \textsc{camb}. In such cases, we exclude these sets from the analysis.}, and for each point in parameter space we compute the CMB temperature power spectrum as described above. 

Here, we do not attempt to also constrain the GP hyperparameters and thus set them to fixed values when training our emulator. In order to determine a suitable value for $\sigma_f$, in our fiducial analysis, we choose it such that the generated primordial power spectra (PPSa) cover a broad range of variations around the \textit{Planck} best-fit power law \( \mathcal{P}_{\text{P18}} \). Specifically, we set it to encompass deviations of around $30-40\sigma$, where $\sigma$ denotes the \textit{Planck} 2018 uncertainty on $A_s$~\cite{2020Planck2018Cosmoparams}. As we are imposing the correlations between the nodes in logarithmic space, this amounts to setting $\sigma_f=0.39$\footnote{By comparing the spread of uncertainties around the mean function in logarithmic space with those in linear space, we ensure that this choice appropriately captures the expected deviations from \( \mathcal{P}_{\text{P18}}\).}. In our fiducial analysis, we furthermore set the length scale hyperparameter to \(l = 0.95\). We find that the rapidly varying PPSa obtained for low correlations lengths can cause the NN training to fail due to more features or ``spikes'' in the generated data. Therefore, we set $l$ to the lowest possible value that still allows our NN to learn efficiently.

Fixing the hyperparameters of the GP can potentially lead to biases in the reconstructed PPS. As per our discussion of the GP mean function, this is of particular concern in those regions of $k$-space where data is sparse (see e.g. Ref.~\cite{2022MNRAS.512.1967R} for a related discussion). As we intentionally restrict our reconstruction to those scales fully covered by the CMB data, this is less of a concern, but we nevertheless test the robustness of our analysis to these hyperparameter choices. Specifically, we run a suite of different emulators with different hyperparameter settings and compare their performance in Sec.~\ref{sec:results} and Appendix \ref{appendix:n(p(k))}. It will be interesting to extend the framework presented in this analysis to also constrain GP hyperparameters alongside the cosmological parameters and PPS nodes, but we leave an investigation thereof to future work.

\subsubsection{Neural network training and evaluation} \label{sec:nn_eval}
We evaluate the performance of our emulator training using a loss curve, and select the model with the minimum validation loss. For further details on the architecture of our emulators and the training procedure, the reader is referred to Appendix \ref{appendix:nn_train}. 

As we are essentially training an emulator to predict CMB temperature power spectra for free-form PPSa, it is important to make sure that our emulators can capture the complex features arising in such an analysis. To this end, we generate a designated test set, which we produce in the same manner as the training set and which consists of approximately \(6 \times 10^3\) samples. We then evaluate the performance of our emulators by comparing the prediction accuracy to the statistical uncertainties for two experiments: following Ref.~\cite{2022CP}, we first compare to the expected instrumental noise and cosmic variance from the upcoming Simons Observatory (SO) \cite{SO}. In addition, we also compare against the statistical errors obtained in the \textit{Planck} 2018 analysis \cite{2019arXiv190712875P}. 
\begin{figure*}[tbp]
  \centering
  \begin{subfigure}{0.49\textwidth}
    \centering
    \includegraphics[scale=0.35]{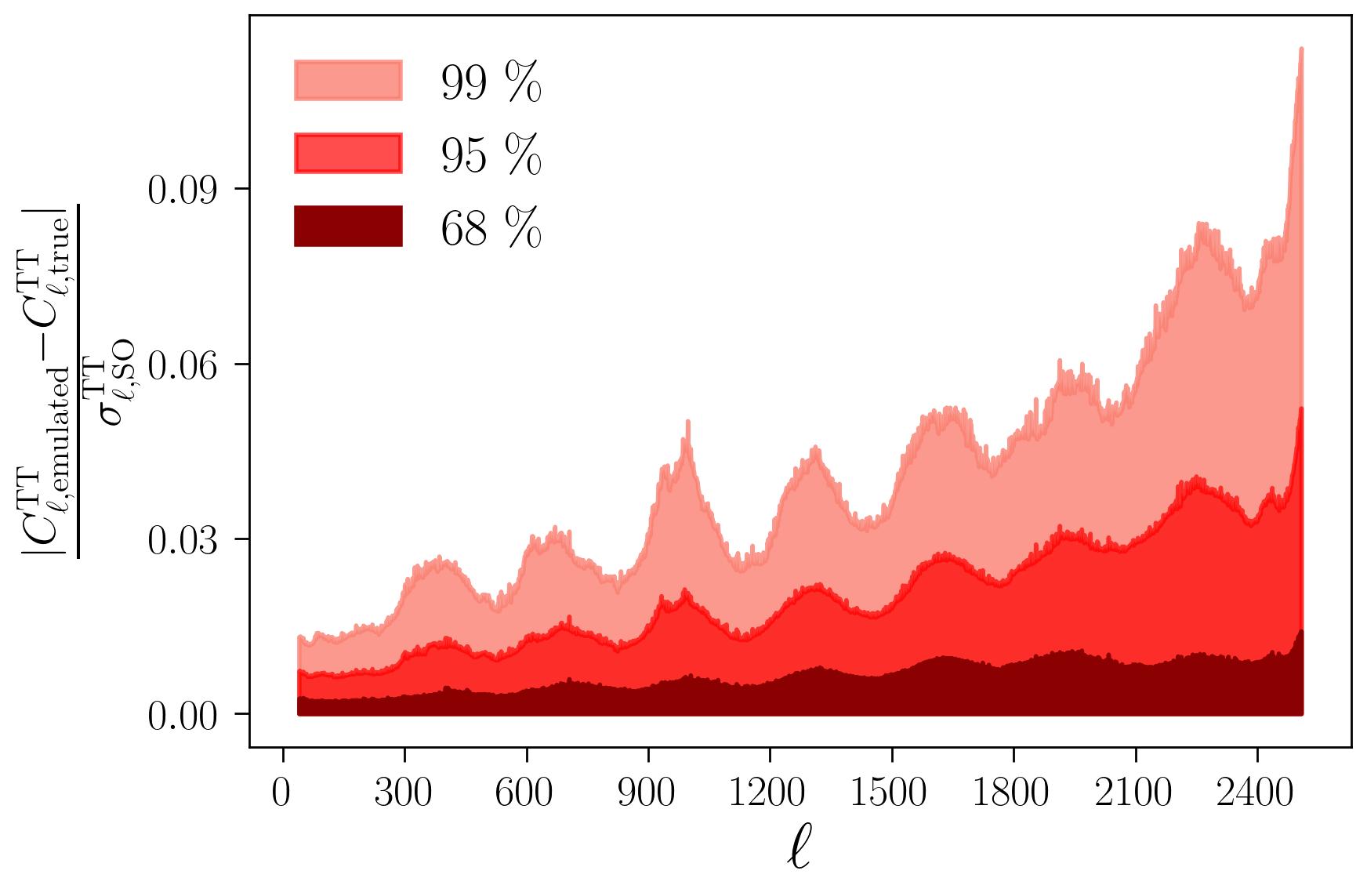}
    \caption[]{}
    \label{fig:train2a}
  \end{subfigure}
  \hfill
  \begin{subfigure}{0.49\textwidth}
    \centering
    \includegraphics[scale=0.35]{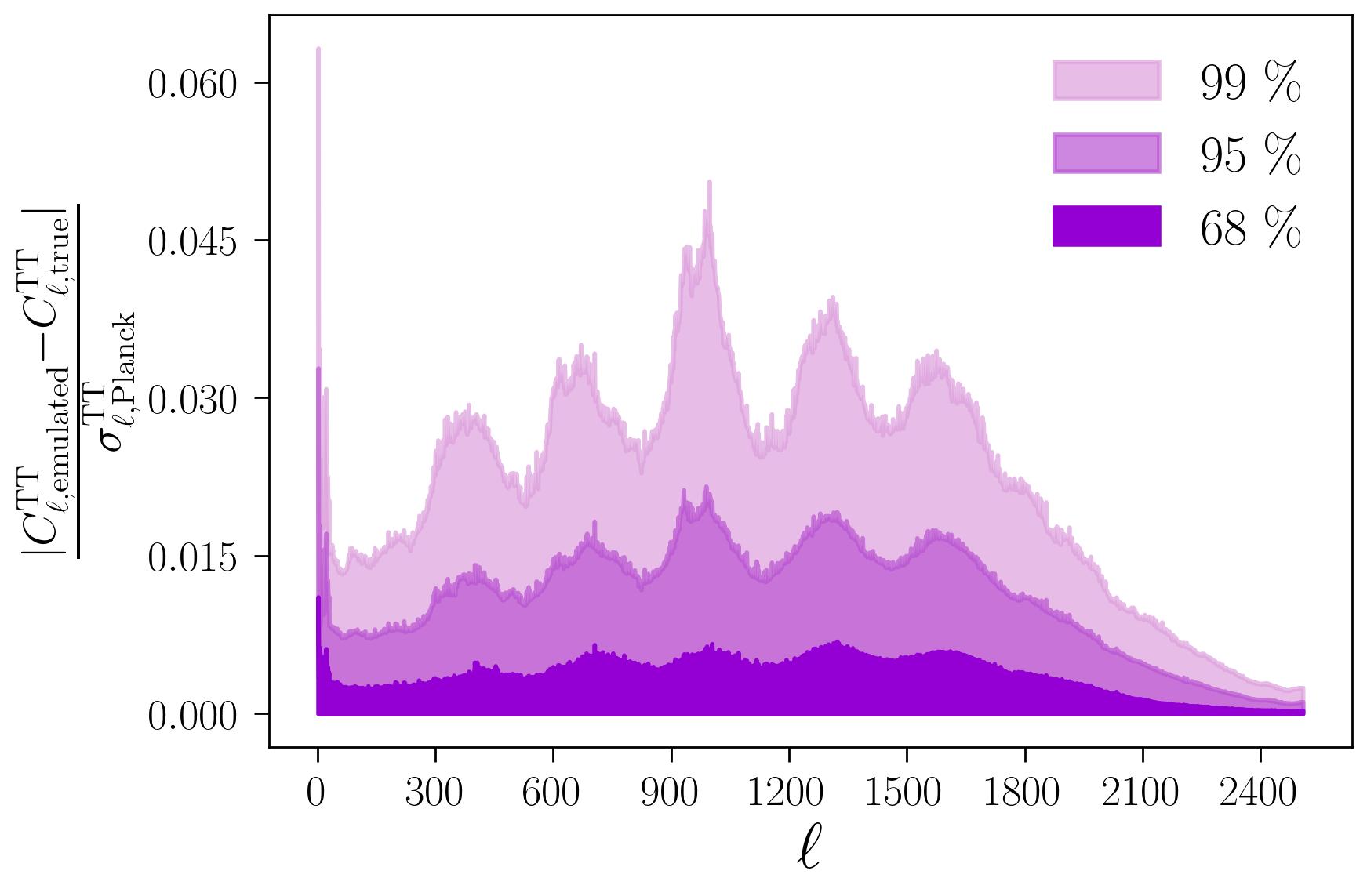}
    \caption[]{}
    \label{fig:train2b}
  \end{subfigure}
  \caption{Emulation accuracy of the predicted CMB temperature power spectra compared with (a) expected Simons Observatory uncertainties across the multipole range used in the analysis and (b) \textit{Planck} 2018 error bars. The different colours indicate regions encompassing the 68th, 95th, and 99th percentiles of the test set.}
  \label{fig:train_2}
\end{figure*}

Following \cite{2022CP}, the emulation error can be computed as
\begin{equation}
    \frac{\left| C_{\ell,\text{emulated}}^{\mathrm{TT}}-C_{\ell,\text{true}}^{\mathrm{TT}}  \right|}{\ \sigma_{\ell}^{\mathrm{TT}}}\;,
\end{equation}
where $C_{\ell,\text{emulated}}^{\mathrm{TT}}$ is the emulated CMB temperature power spectrum and $C_{\ell,\text{true}}^{\mathrm{TT}}$ is the true power spectrum generated using \textsc{camb}. The quantity $\sigma_{\ell}^{\mathrm{TT}}$ denotes the experimental uncertainty. For SO, we follow Ref.~\cite{2022CP} and set
\begin{equation}
    \sigma_{\ell,\text{SO}}^{\mathrm{TT}} =\sqrt{\frac{2}{f_{\text{sky}}(2\ell +1)}} \left( C_{\ell,\text{Planck}}^{\mathrm{TT}} + N_\ell^{\mathrm{TT}} \right)\;.
\end{equation}
Here, $f_\text{sky} = 0.4$ represents the observed sky fraction, $C_{\ell,\text{Planck}}^{\mathrm{TT}}$ represents a reference power spectrum calculated with \textit{Planck} 2018 best-fit parameters using \textsc{camb}, while $N_\ell^{\mathrm{TT}}$ denotes the goal noise curves that account for instrumental and atmospheric effects, as specified for the Simons Observatory \cite{SO}\footnote{For large angular scales that are not probed by SO, it incorporates Planck data over a reduced sky fraction to model the co-added noise curves~\cite{SO}.}. For \textit{Planck} we use $\sigma_{\ell,\text{Planck}}^{\mathrm{TT}}$ which represents the error bars from the \textit{Planck} 2018 CMB temperature anisotropy data \cite{2019arXiv190712875P,2020Planck2018Cosmoparams}. The results of this analysis can be seen in Fig.~\ref{fig:train_2} where we show the emulation error as a function of angular multipole $\ell$ for both cases. As can be seen, nearly 99\% of the obtained deviations fall within less than \(0.09\sigma_{\ell, \rm{SO}}^{\rm{TT}}\) and \(0.06\sigma_{\ell, \rm{Planck}}^{\rm{TT}}\) respectively, thus demonstrating the reliability of the emulator and its ability to reproduce the complex power spectra required for our analysis. 

\begin{figure*}[tbp]
  \centering
  \begin{subfigure}{0.48\textwidth}
    \centering
    \includegraphics[scale=0.30]{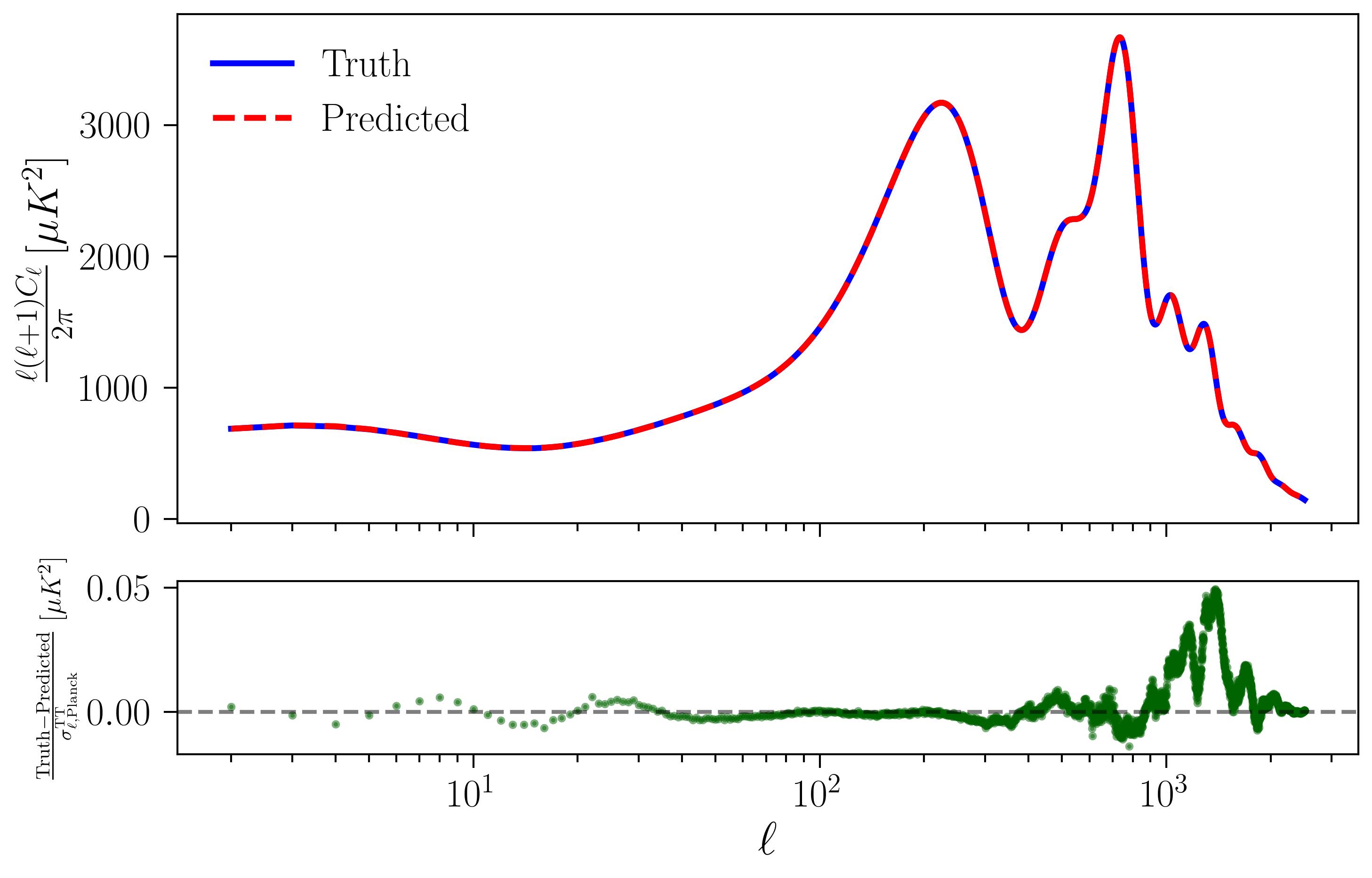}
    \caption[]{}
    \label{fig:train1a}
  \end{subfigure}
  \hfill
  \begin{subfigure}{0.48\textwidth}
    \centering
    \includegraphics[scale=0.30]{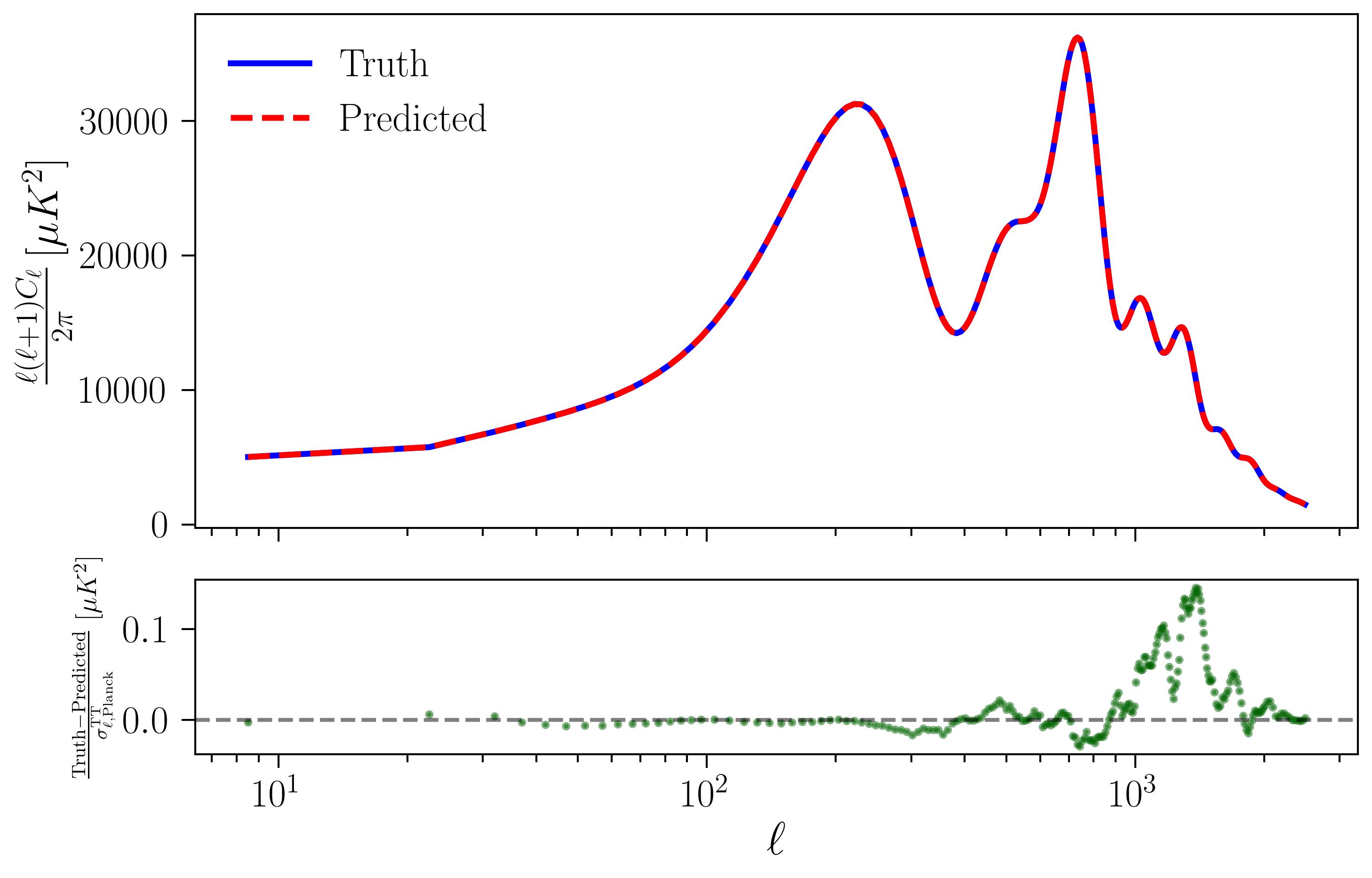}
    \caption[]{}
    \label{fig:train1b}
  \end{subfigure}
  \caption{Comparison between the emulated and ground-truth CMB temperature power spectra for a randomly selected test sample. Panel (a) shows the unbinned
  spectrum, while panel (b) shows the binned version. In both cases, the lower sub-panels display the residuals, computed as the difference between the ground-truth and the emulator prediction, normalized by the corresponding Planck 2018 uncertainties across the full multipole range.}
  \label{fig:train_1}
\end{figure*}

To further illustrate the emulator’s performance, in Fig.~\ref{fig:train_1} we show a comparison between an arbitrary CMB temperature power spectrum from the test set and its corresponding emulator prediction. The left panel displays the unbinned spectrum, while the right panel shows the binned version using the same binning scheme employed in the likelihood evaluations. In both cases, the bottom sub-panels present the residuals, normalized by the \textit{Planck} 2018 uncertainties. Across the entire multipole range, the deviations remain well below $1\sigma$, despite the non-trivial features exhibited by the CMB power spectrum\footnote{These features are entirely determined by corresponding structures in the PPS.}. Although some residual structure persists in the binned version, its amplitude remains consistently below the \textit{Planck} uncertainties in all bins. This suggests that any systematic component in the residuals is too small to introduce a significant bias in the reconstructed bandpowers. Given the large number of parameters involved in the emulation, this robustness is especially crucial to
ensure accurate predictions across the parameter space.

\section{Data and Likelihood} \label{sec:data_likelihood}

To reconstruct the PPS, we employ CMB temperature power spectra from the third data release of \textit{Planck} (PR3)\footnote{The data is obtained from \textit{Planck} Legacy Archive: \url{https://pla.esac.esa.int}}\cite{planckoverview2018,2019arXiv190712875P,2020Planck2018Cosmoparams}. Following Ref.~\cite{2019arXiv190712875P}, we employ two different likelihoods for the large-scale and small-scale CMB temperature power spectrum.

For the high-$\ell$ ($\ell\geq 30$) part of the CMB likelihood, we follow the \textsc{Python}  implementation of \textit{Planck}'s \texttt{plik\_lite} likelihood \cite{2019arXiv190712875P}, \texttt{Planck-lite-py}\footnote{The code is available at \url{https://github.com/heatherprince/planck-lite-py.git}.}, which is described in Ref.~\cite{heatherplancklite}. This likelihood is pre-marginalised over \textit{Planck} high-$\ell$ CMB temperature foreground parameters, but has been shown to be consistent with the full likelihood in Ref.~\cite{heatherplancklite}. The only remaining nuisance parameters in \texttt{Planck-lite-py} is the overall multiplicative \textit{Planck} calibration factor, $A_{\text{Planck}}$. 

At low multipoles, \(2 < \ell < 30\), the likelihood of the angular power spectrum is non-Gaussian. While the official \textit{Planck} analysis employs a hybrid, map-based approach \cite{2019arXiv190712875P}, Ref.~\cite{heatherplancklite} showed that the low-$\ell$ CMB temperature power spectrum can be efficiently compressed into two Gaussian-distributed data points. In our analysis, we employ the \textit{Planck} 2018 version of the compressed low-$\ell$ likelihood presented in Ref.~\cite{heatherplancklite}\footnote{The low-$\ell$-compressed likelihood, as two Gaussian bins, is implemented in the same code as an option to be included alongside the high-$\ell$ likelihood via a boolean flag.}, which has been shown to give results consistent with the official analysis.  

To perform cosmological inference from the \textit{Planck} low- and high-$\ell$ likelihoods, we employ two complementary sampling algorithms and associated packages. The first approach uses Importance Nested Sampling (\textsc{INS}) \cite{2019OJAp....2E..10F} as implemented in the \textsc{Nautilus} \cite{nautilus2023MNRAS.525.3181L} package\footnote{\url{https://github.com/johannesulf/nautilus.git}}.

Nested Sampling \cite{skilling2004nested} is an algorithm to generate samples from posterior distributions and compute Bayesian evidences. It starts with a set of ``live points'' drawn from the prior distribution. At each iteration, the point with the lowest likelihood is replaced by a new point with a higher likelihood, progressively focusing on high-likelihood regions. This process continues until further contributions to the evidence are negligible. The prior volume represented by the live set decreases exponentially and the weighted posterior distribution is reconstructed using both the inactive and remaining live points.

\textsc{INS} enhances standard nested sampling by making better use of all likelihood evaluations. Instead of assigning posterior sample weights based solely on the shrinking-volume assumption, INS estimates the underlying sampling distribution and uses this to assign importance weights to all sampled points, including those outside the live set. This leads to a more accurate reconstruction of both the posterior and the Bayesian evidence \cite{2019OJAp....2E..10F,nautilus2023MNRAS.525.3181L}. \textsc{Nautilus} further advances this methodology by incorporating NN regression to refine proposal distributions, concentrating sampling in regions of high posterior probability while ensuring thorough exploration of the entire parameter space. Additionally, it allows for drawing extra samples after the initial run, increasing the effective sample size and further improving the accuracy of the estimated posterior~\cite{nautilus2023MNRAS.525.3181L}.

As a consistency test, we compare the results from \textsc{Nautilus} with those from \textsc{emcee}, a standard MCMC sampler \cite{emcee2013PASP..125..306F}. To simplify the comparison, we fix the 20 GP nodes to the \textit{Planck} 2018 best-fit mean function (see Eq.~\ref{eq:mean_func}) and sample only the 4 cosmological parameters and the nuisance parameter $A_{\text{Planck}}$. We find that both methods yield consistent results.

A key challenge with stochastic sampling algorithms such as traditional MCMC and \textsc{INS} is that they can become very inefficient in exploring complex, high-dimensional posterior distributions. This is mainly due to the fact that finding the high-posterior regions in these cases becomes increasingly difficult. In order to overcome this challenge, gradient-based samplers, such as HMC, leverage the gradient of the posterior distribution to guide the sampling process, resulting in a more efficient exploration of parameter space \cite{DUANE1987216,HMC,2011hmcm.book..113N}. HMC simulates Hamiltonian trajectories through parameter space to efficiently propose new points, which can significantly speed up the exploration of high-dimensional parameter spaces. However, the performance of HMC relies on two key parameters, the step size and the number of steps employed when simulating the Hamiltonian trajectories, which thus need to be adapted to the problem at hand. The No-U-Turn Sampler (\textsc{NUTS}) \cite{2011arXiv1111.4246H} is an extension of HMC, which dynamically tunes these parameters. The essential idea underlying \textsc{NUTS} is to prevent the Hamiltonian trajectories to turn back on themselves by avoiding U-turns in parameter space.

\begin{table}[tbp]
\centering
\renewcommand{\arraystretch}{1.1} 
\setlength{\tabcolsep}{12pt}
\begin{tabular}{c|c|c}
\toprule
\textbf{Parameter}    & \textbf{Prior Distribution}                & \textbf{Best Fit Value} \\
\midrule
$\omega_\mathrm{b}$            & $\mathcal{U}[0.015,\, 0.03]$                   & $0.02191 \pm 0.00039$   \\
$\omega_\mathrm{cdm}$        & $\mathcal{U}[0.05,\, 0.3]$                    & $0.1218 \pm 0.0029$       \\
$H_{0}$               & $\mathcal{U}[50,\, 90]$                       & $66.26 \pm 1.35$         \\
$\tau$                & $\mathcal{N}(0.067,\, 0.023)$      & $0.0745 \pm 0.0221$     \\
$A_{\text{Planck}}$   & $\mathcal{N}(1,\, 0.0025)$           & $1.001 \pm 0.002$       \\
GP Nodes              & $\mathcal{N}(0,\, 1)$              & See Fig.~\ref{fig:nuts_ns1}\\
\bottomrule
\end{tabular}
\caption{\label{tab:sampling_tab} Cosmological and nuisance parameters varied in this work, alongside their associated priors and posterior means obtained in our fiducial analysis. The uncertainties denote the 68\% C.L. To improve readability, we do not show the recovered GP nodes in the table.}
\end{table}

Given the high-dimensional parameter spaces involved in our analysis, we additionally employ HMC sampling methods. Since HMC relies on gradient information, it necessitates a fully differentiable analysis pipeline. To this end, we implement the CMB temperature likelihoods using the automatically differentiable \texttt{JAX} library \cite{jax2018github, Frostig2018CompilingML}. These likelihoods are interfaced with our NN emulators through the \texttt{Cosmopower-JAX} framework\footnote{\url{https://github.com/dpiras/cosmopower-jax.git}} to ensure that our theoretical predictions remain compatible with \texttt{JAX} \cite{CPJ2023OJAp....6E..20P}. We then use the \textsc{NUTS} variant of HMC as implemented in the \texttt{NumPyro} library\footnote{\url{https://github.com/pyro-ppl/numpyro.git}} \cite{bingham2019pyro, phan2019composable} to obtain joint constraints on cosmological and PPS parameters.

\begin{figure*}[tbp]
\centering % \begin{center}/\end{center} takes some additional vertical space
\includegraphics[width=1.0\textwidth]{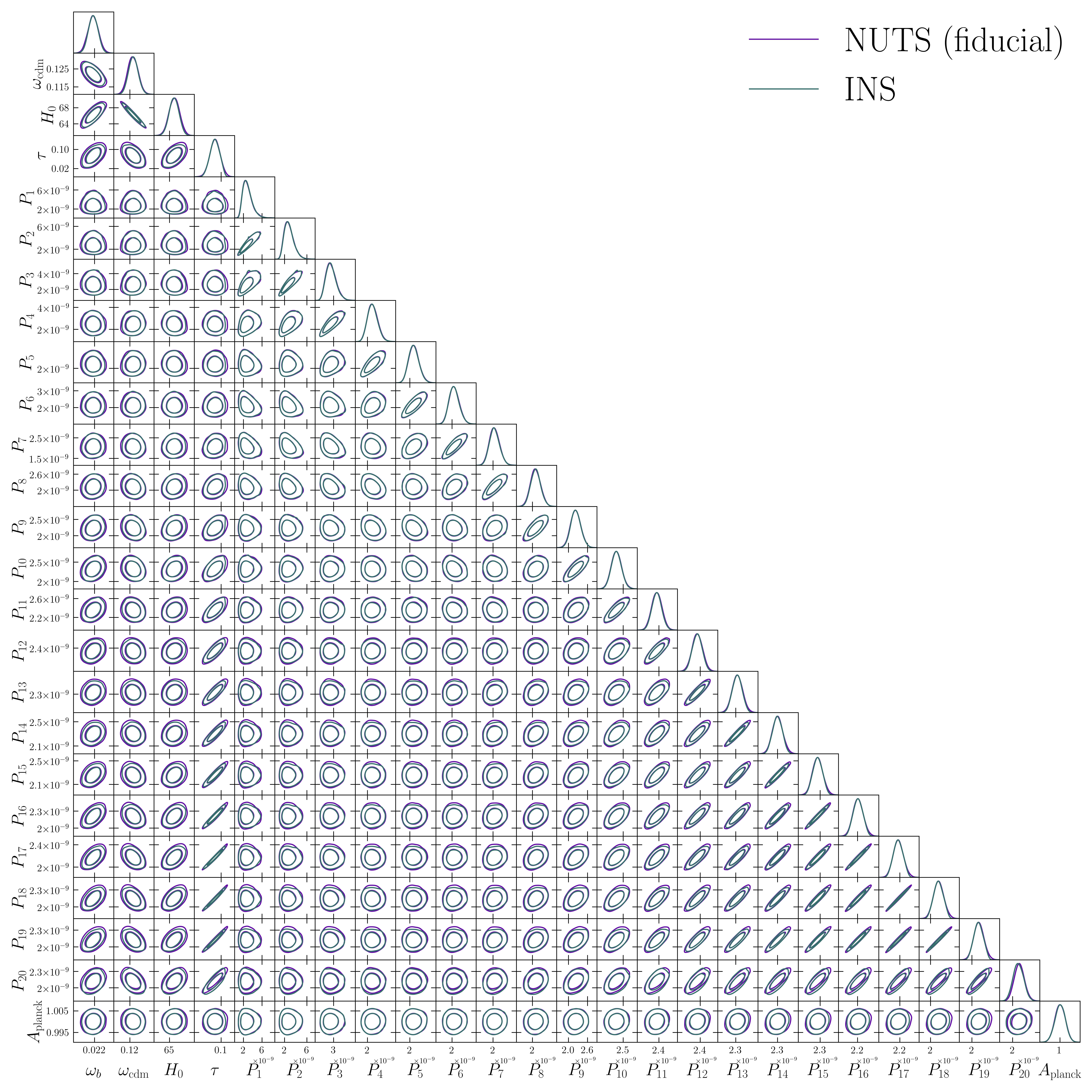}
\caption{\label{fig:nuts_ns1} Constraints on the cosmological parameters, the PPS values calculated at each mode using the constrained GP nodes and the nuisance parameter obtained in our fiducial analysis with \textsc{NUTS} and \textsc{INS}. The contours represent the 68\% and 95\% confidence intervals, with 1-D marginals for each parameter.}
\end{figure*}

The priors used in our analysis are summarized in Tab.~\ref{tab:sampling_tab}; these priors are mostly flat and match the ranges used for our emulators (see Tab.~\ref{tab:prior_tab}), with two exceptions. First, we follow Ref.~\cite{2019arXiv190712875P} and impose a Gaussian prior on the \textit{Planck} calibration parameter \(A_{\text{Planck}}\). Second, as we do not include polarisation information in our analysis, we need to incorporate it in the form of a prior on the optical depth to reionization, $\tau$, as otherwise we would not be able to constrain the amplitude of the PPS due to its degeneracy with $\tau$. In our fiducial analysis, we impose a Gaussian prior on $\tau$ that compresses the \textit{Planck} 2015 low-$\ell$ polarisation data \cite{planck2015likelihood}, following Ref.~\cite{heatherplancklite}. Although this prior is based on an earlier analysis than the updated \textit{Planck} 2018 results~\cite{2019arXiv190712875P}, it still encompasses the more recent best-fit value. This makes it a less informative but more flexible choice, allowing the data to play a greater role in informing the reconstruction. Additionally, low-$\ell$ polarization based estimates of $\tau$ typically include low-$\ell$ temperature information as well. Thus, using the wider 2015 prior helps minimize the risk of double-counting this temperature information. 

We note that the \textit{Planck} 2015 $\tau$ measurements are known to be biased due to systematics, such as calibration leakage and instrumental effects, which were mitigated in the \textit{Planck} 2018 results~\cite{2019arXiv190712875P}. However, due to the broadness of our adopted 2015 $\tau$ prior, this choice does not bias our results, as we demonstrate in Sec.~\ref{sec:results}, where we explicitly test how our constraints change when using the higher-precision $\tau$ measurements from the \textit{Planck} 2018 results~\cite{2019arXiv190712875P}. We therefore keep the reconstruction based on \textit{Planck} 2015 $\tau$ measurements as our fiducial baseline and leave a full joint analysis of CMB temperature and polarisation to future work.

\section{Results} \label{sec:results}
Using our differentiable pipeline, we can now move on to reconstructing the PPS using \textit{Planck} 2018 CMB temperature anisotropy data. 
\begin{figure*}[tbp]
\centering % \begin{center}/\end{center} takes some additional vertical space
\includegraphics[width=0.63\textwidth]{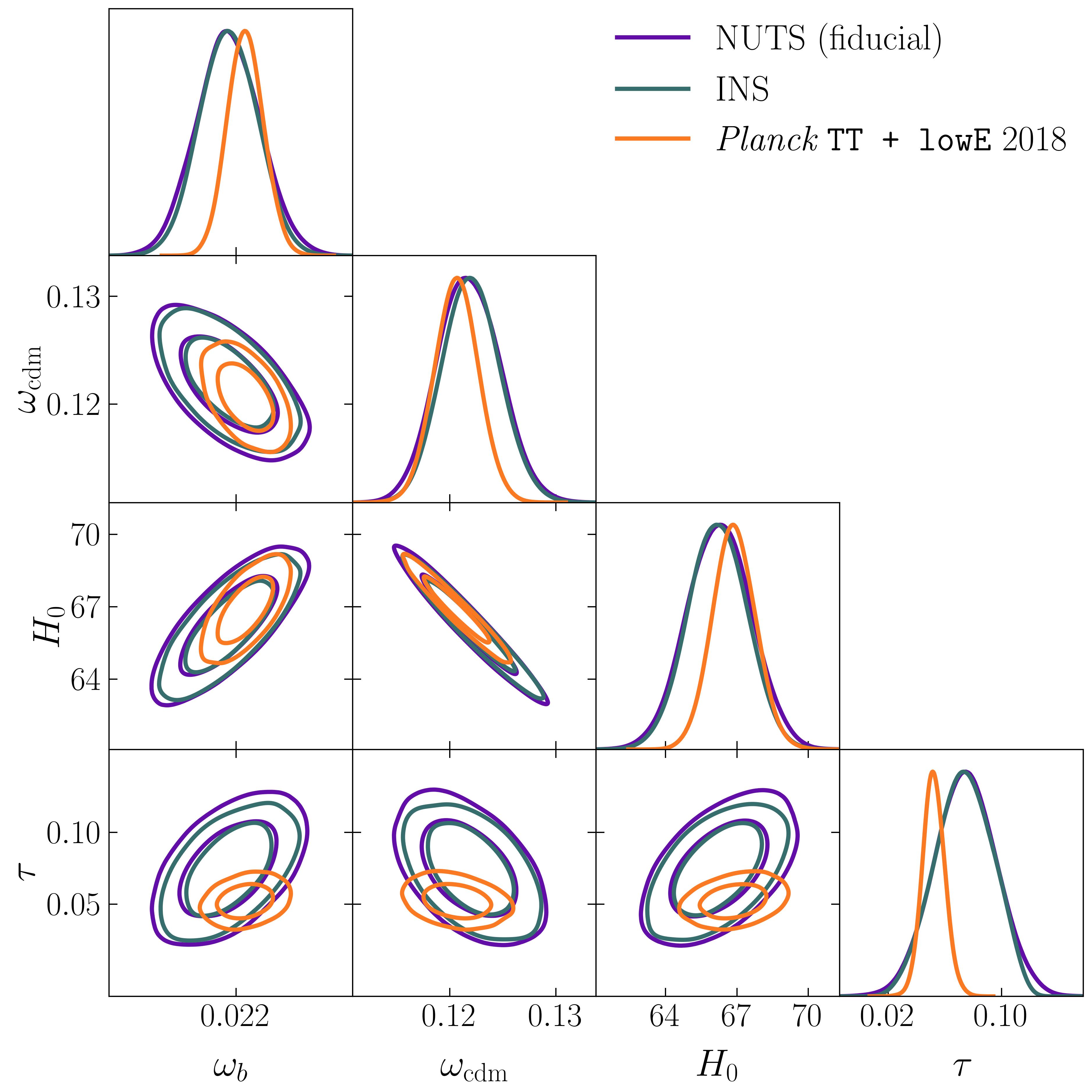}
\caption{\label{fig:cosmo_params} Constraints on the cosmological parameters from our fiducial analysis compared to the \textit{Planck} 2018 $\texttt{TT+lowE}$ results. The PPS and nuisance parameter have been marginalized over. The contours represent the 68\% and 95\% confidence intervals, with 1-D marginals for each parameter.}
\end{figure*}
In Fig.~\ref{fig:nuts_ns1}, we show the joint constraints on the PPS values obtained at each $k$ mode, alongside the cosmological and nuisance parameters, as obtained in our fiducial analysis both for \textsc{INS} and \textsc{NUTS}. Figure \ref{fig:cosmo_params} illustrates the same constraints but focuses on the cosmological parameters, and compares our results to those obtained from \textit{Planck} 2018. Focusing first on the results obtained using our fiducial sampling algorithm \textsc{NUTS}, we can see from Figs.~\ref{fig:nuts_ns1}, \ref{fig:cosmo_params} and Tab.~\ref{tab:sampling_tab} that the inferred constraints on the cosmological parameters $\omega_{b}$, $\omega_{\mathrm{cdm}}$, and $H_0$, agree with the official \textit{Planck} 2018 results to within 1$\sigma$ \cite{2020Planck2018Cosmoparams}, as expected. 
Specifically, we compare to the \texttt{TT+lowE} constraints listed in Table 2, column 2 of Ref.~\cite{2020Planck2018Cosmoparams}, which is the analysis variant closest to the data combination considered in this work. The difference in our inferred value of $\tau = 0.0745 \pm 0.0221 $ and the value of $\tau = 0.0522 \pm 0.0080$ reported by the \textit{Planck} Collaboration in their 2018 results is due to our use of a wider prior based on the \textit{Planck} 2015 results and Ref.~\cite{heatherplancklite}. The freedom allowed by our prior choice pushes our fiducial constraints to higher values for the optical depth as compared to the results from \textit{Planck} 2018 as we describe below.

\begin{figure*}[tbp]
\centering % \begin{center}/\end{center} takes some additional vertical space
\includegraphics[width=0.82\textwidth]{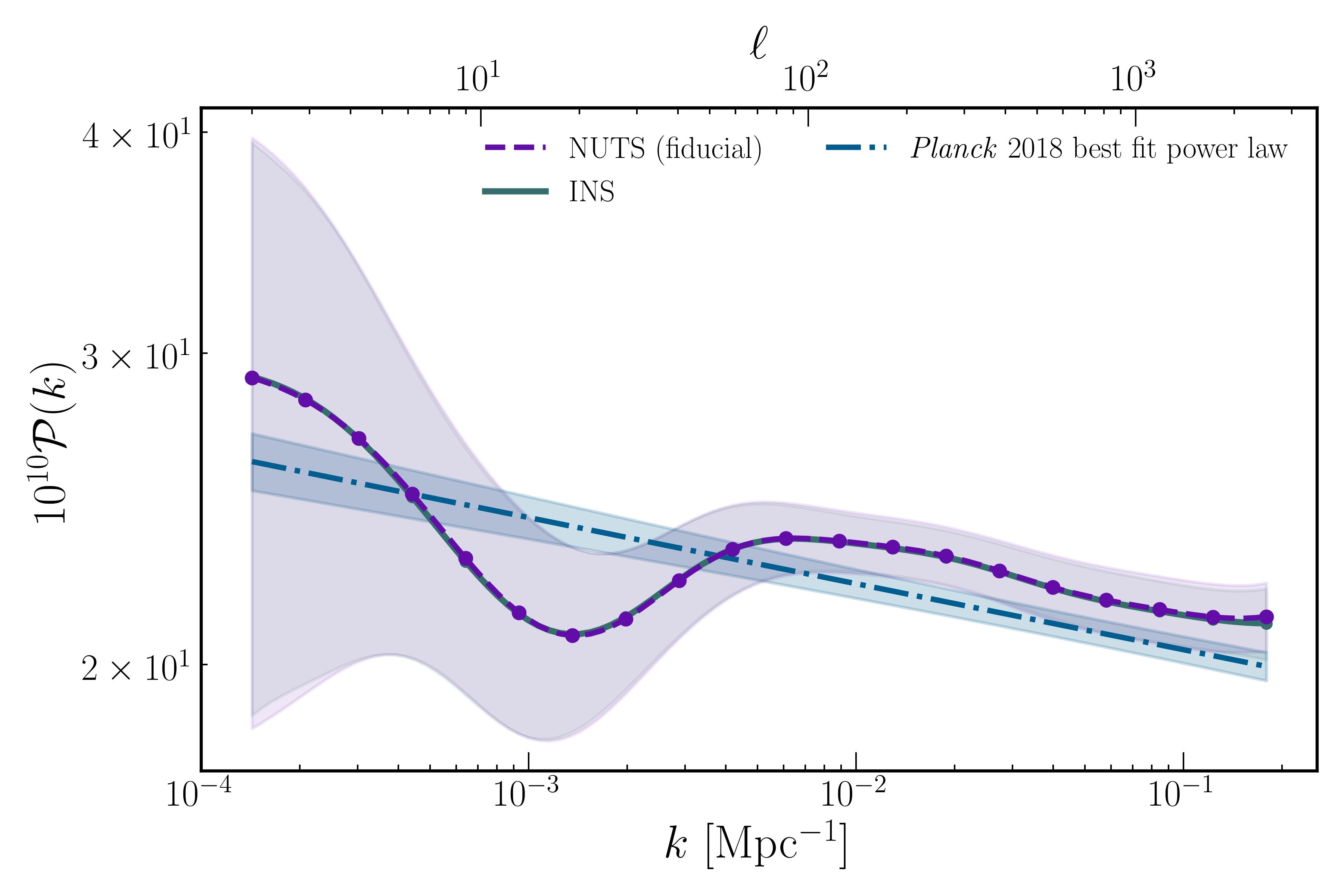}
\caption{\label{figure:nuts_ns_recons}The mean reconstructed PPS for our fiducial setup obtained using \textsc{INS} and \textsc{NUTS}. The shaded regions denote the 68\% confidence intervals, and the points highlight the GP nodes used in our analysis.}
\end{figure*}

In Fig. \ref{figure:nuts_ns_recons}, we show the corresponding reconstructed PPS, alongside the best-fit power-law spectrum from \textit{Planck} 2018. We compute the power-law using Eq.~\ref{eq:powerlaw}, with $n_s$ and $A_s$ values taken from the \textit{Planck} 2018 \texttt{TT+lowE} results from Table 2 in Ref.~\cite{2020Planck2018Cosmoparams}. As a visual illustration of the uncertainties on the PPS obtained from the official \textit{Planck} 2018 analysis, the shaded region around the \textit{Planck} power-law denotes variations allowed by the  1$\sigma$ uncertainty on $A_s$ and $n_s$ as derived in Ref.~\cite{2020Planck2018Cosmoparams}. For our PPS reconstruction, the solid and dashed line shows the mean across all samples, and the shaded region encompasses the $68\%$ confidence interval. As can be seen, the uncertainties on the reconstructed PPS at small $k$ are quite large. However, our results are in good agreement with the \textit{Planck} power law. Despite the low signal-to-noise, our reconstructed PPS shows a hint of a dip at a scale of around $k \sim 1.57 \times 10^{-3}$ $\text{Mpc}^{-1}$. This corresponds to angular multipoles in the range $\ell \sim 15-30$, where all CMB analyses so far have reported a dip in the CMB temperature power spectrum (see e.g. Ref.~\cite{2003ApJS..148....1B, 2019arXiv190712875P}). Our results suggest that this observed lack of power causes a corresponding dip in our reconstructed PPS, consistent with previous works (see e.g. \cite{2016A&A...594A..20P, 2020A&A...641A..10P}).
The significant uncertainties observed on the largest scales can primarily be attributed to cosmic variance. Additionally, we suspect that our use of the compressed \texttt{Planck-lite-py} low-$\ell$ likelihood, as discussed in Section~\ref{sec:data_likelihood}, may contribute to these uncertainties. While Ref.~\cite{heatherplancklite} shows that the likelihood compression is lossless for a $\Lambda$CDM cosmological model, this might not be the case when allowing a free-form PPS. Focusing on small scales (i.e. high $k$), we recover the nearly scale-invariant behaviour predicted by slow-roll single-field inflation. Comparing to the power-law PPS obtained in \textit{Planck} 2018, while the two results remain consistent at 1$\sigma$, we find our reconstruction to systematically overestimate the PPS amplitude. To evaluate the balance between improved data fit and increased model complexity, we perform a model comparison using the Akaike Information Criterion (AIC)~\cite{1974ITAC...19..716A}. Although our fiducial reconstruction yields a lower $\chi^2$, indicating a better fit to the data, the increased number of free parameters incurs a substantial AIC penalty. Specifically, the fiducial model uses 25 parameters and achieves an AIC of 251.4, while the \textit{Planck} best-fit power-law, with 7 parameters, results in an AIC of 223.2. Consequently, the data does not statistically favor the more complex model over \textit{Planck} power-law according to the AIC criterion.
\begin{figure*}[tbp]
\centering % \begin{center}/\end{center} takes some additional vertical space
\includegraphics[width=0.82\textwidth]{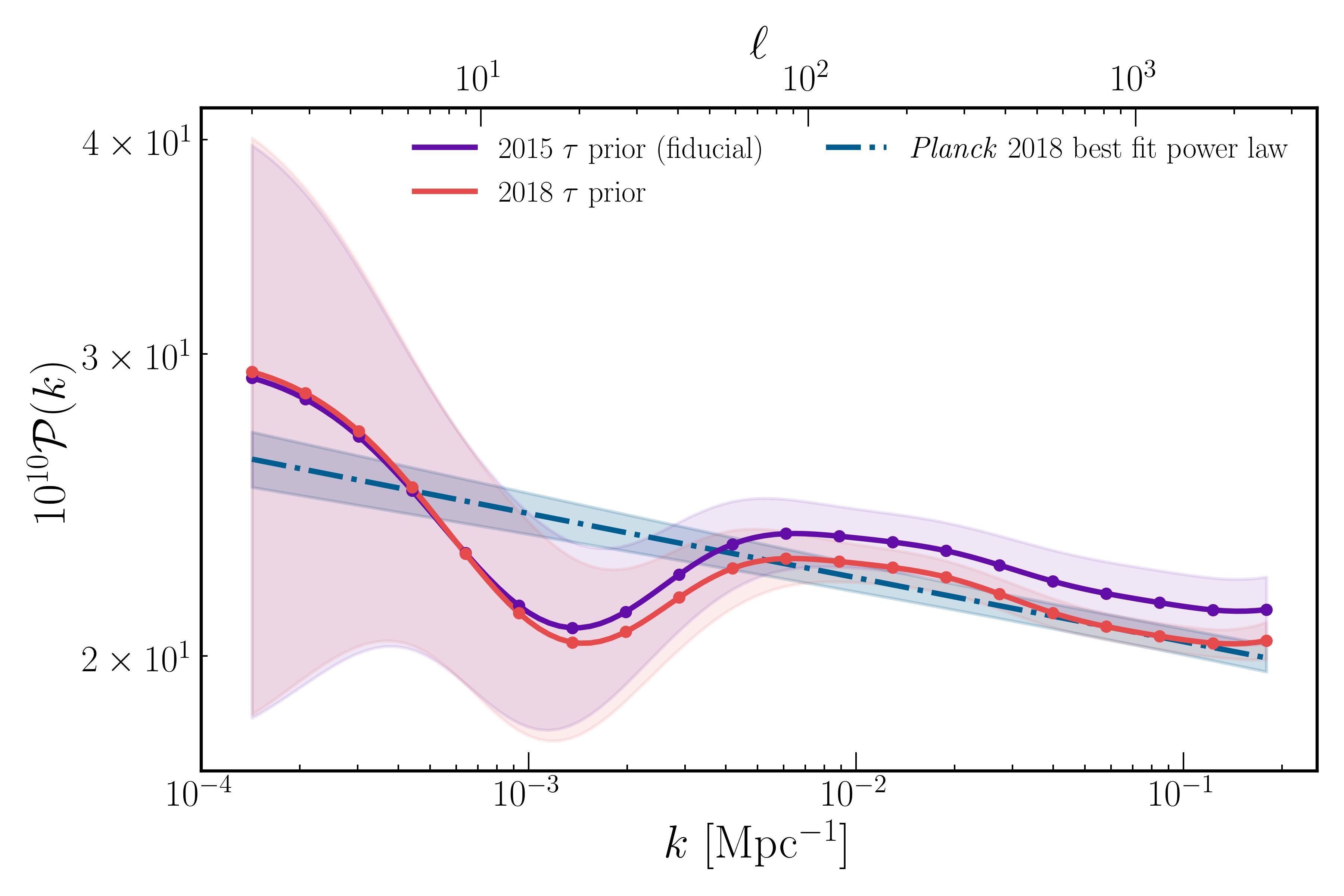}
\caption{\label{fig:diff_tau}Comparison of the mean reconstructed PPS for our fiducial setup using two different $\tau$ priors from \textit{Planck} 2015 and 2018. The shaded regions represent the $68\%$ confidence interval.}
\end{figure*}

In order to better understand this discrepancy in the inferred amplitude of the reconstructed PPS, we repeat our analysis using a $\tau$ prior informed by the low-$\ell$ polarisation likelihood from the updated \textit{Planck} 2018 measurements, specifically from the HFI \cite{PlanckHFI,2019arXiv190712875P}. As compared to the Low-Frequency Instrument based results reported by \textit{Planck} 2015, the HFI measurements result in a significantly lower value of the optical depth, which following Ref.~\cite{2019arXiv190712875P}, we summarize into a Gaussian prior with $\tau=0.0506 \pm  0.0086$. The resulting reconstructed PPS is shown in Fig.~\ref{fig:diff_tau} alongside our fiducial results; the full contour plot is deferred to the Appendix \ref{appendix:sampling}. Adopting the \textit{Planck} 2018 prior reduces the uncertainties on the PPS at small scales, indicating that these uncertainties are primarily driven by the choice of $\tau$ prior. In contrast, the uncertainties on large scales remain largely unchanged, as they are dominated by cosmic variance or the use of the compressed likelihood. Reassuringly, our results using the \textit{Planck} 2018 optical depth prior move closer to the \textit{Planck} 2018 best-fit power law model. This suggests that the overestimation of the PPS amplitude observed in our fiducial analysis is driven by our choice of a wider $\tau$ prior. As discussed above, this leads to an inferred value of $\tau$ significantly higher than the official results from \textit{Planck} 2018. Due to the degeneracy between $\tau$ and $A_s$, this higher optical depth leads to a correspondingly higher PPS amplitude, which we observe in our reconstruction. We note that the $\tau$-$A_s$ degeneracy can also be observed in Fig.~\ref{fig:nuts_ns1} as a degeneracy between the high-$k$ PPS nodes $\mathcal{P}_i$ and the optical depth $\tau$.

\begin{figure*}[tbp]
\centering % \begin{center}/\end{center} takes some additional vertical space
\includegraphics[width=0.82\textwidth]{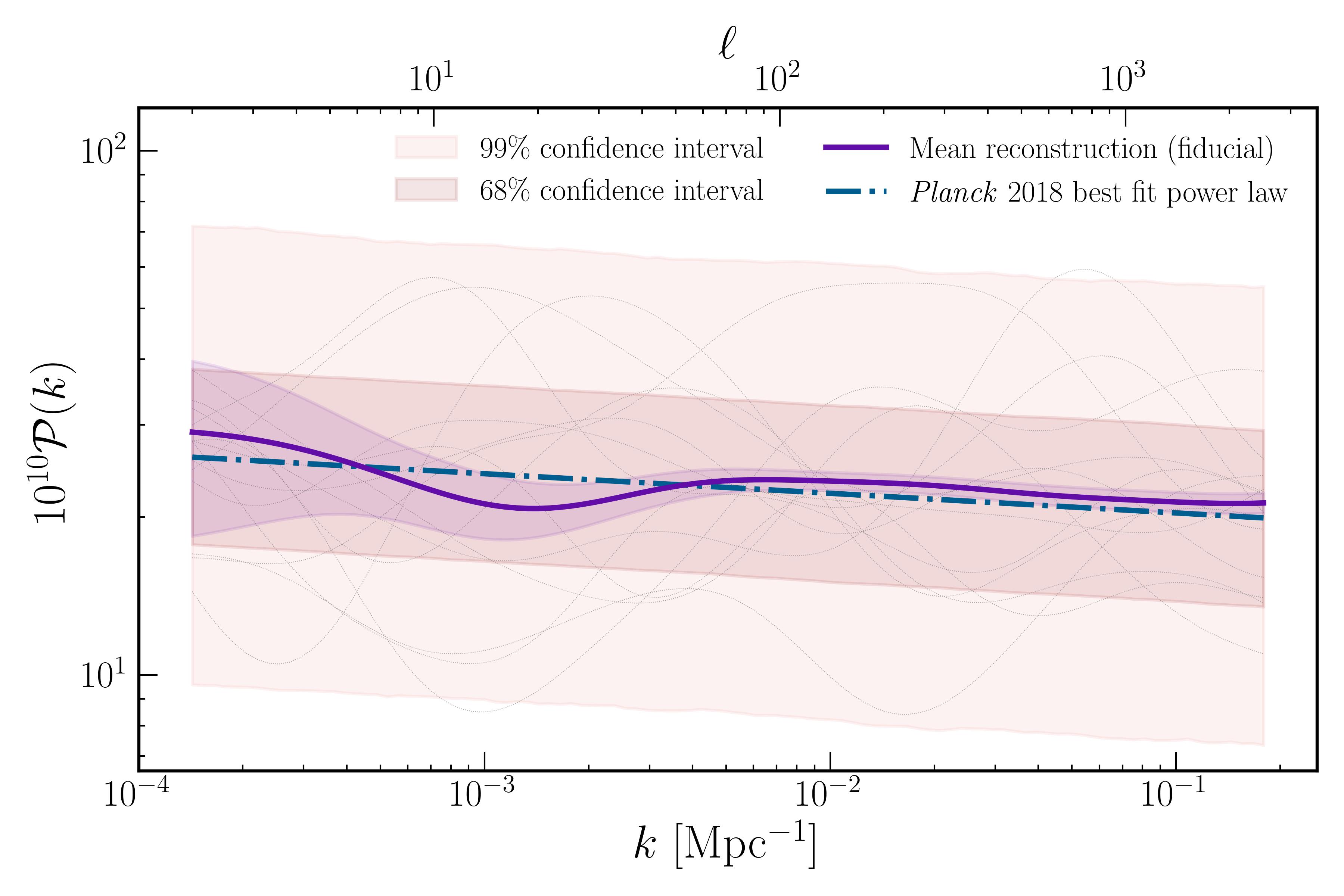}
\caption{\label{figure:gp_prior} Fiducial PPS reconstruction compared to the prior imposed on the GP (see Sec.~\ref{sec:data_gen} for details on the GP prior). The shaded region around the \textit{Planck} best-fit power law illustrates the $68\%$ (dark) and $99\%$ (light) percentile of the GP prior, while the gray dotted lines show random draws from this prior. The shaded region around the reconstruction represents the $68\%$ confidence interval.}
\end{figure*}

To test the robustness of our HMC pipeline, we compare the fiducial constraints obtained using \textsc{NUTS} with those derived from \textsc{INS}. As shown in Figs. \ref{fig:nuts_ns1} and \ref{fig:cosmo_params}, the posterior means from both \textsc{NUTS} and \textsc{INS} are consistent. This agreement is also supported by the mean reconstructed PPS in Fig.~\ref{figure:nuts_ns_recons}. 
In order to further assess the robustness of both the \textsc{INS}- and \textsc{NUTS}-based results, we perform a series of consistency checks by varying key hyperparameters of each algorithm. The details of these tests are provided in Appendix~\ref{appendix:sampling}. We find that our constraints remain stable under these variations. 

Since the reconstructed PPS from both methods is consistent, we adopt gradient-based sampling using \textsc{NUTS} for the remainder of this analysis. This choice is also driven by the fact that \textsc{NUTS} significantly accelerates the inference process compared to \textsc{INS}, with \textsc{NUTS} taking 2 hours to obtain the constraints shown in Fig.~\ref{fig:nuts_ns1} compared to around 2 days it took with \textsc{INS}. Details of these runtime comparisons and hardware used are included in Appendix~\ref{appendix:sampling}. 

Finally, in Fig.~\ref{figure:gp_prior}, we show the comparison of our fiducial PPS reconstruction to the GP prior employed in our analysis. We show both the $68\%$ and the $99\%$ interval of random realizations drawn from the GP prior. As can be seen, our constraints on the PPS are prior-dominated at the largest scales (up to $k_\mathrm{min}\sim 3\times 10^{-4}\ \text{Mpc}^{-1}$), which we attribute to cosmic variance and our use of a compressed low-$\ell$ likelihood. On smaller scales on the other hand, the likelihood dominates over the prior, showing that both the lack of power observed at around $k \sim 1.57 \times 10^{-3}\ \text{Mpc}^{-1}$ and the near scale-invariant behaviour at small scales are robust to our prior choice (we will come back to this when discussing GP hyperparameter choices below). 

\begin{figure*}[tbp]
\centering % \begin{center}/\end{center} takes some additional vertical space
\includegraphics[width=0.82\textwidth]{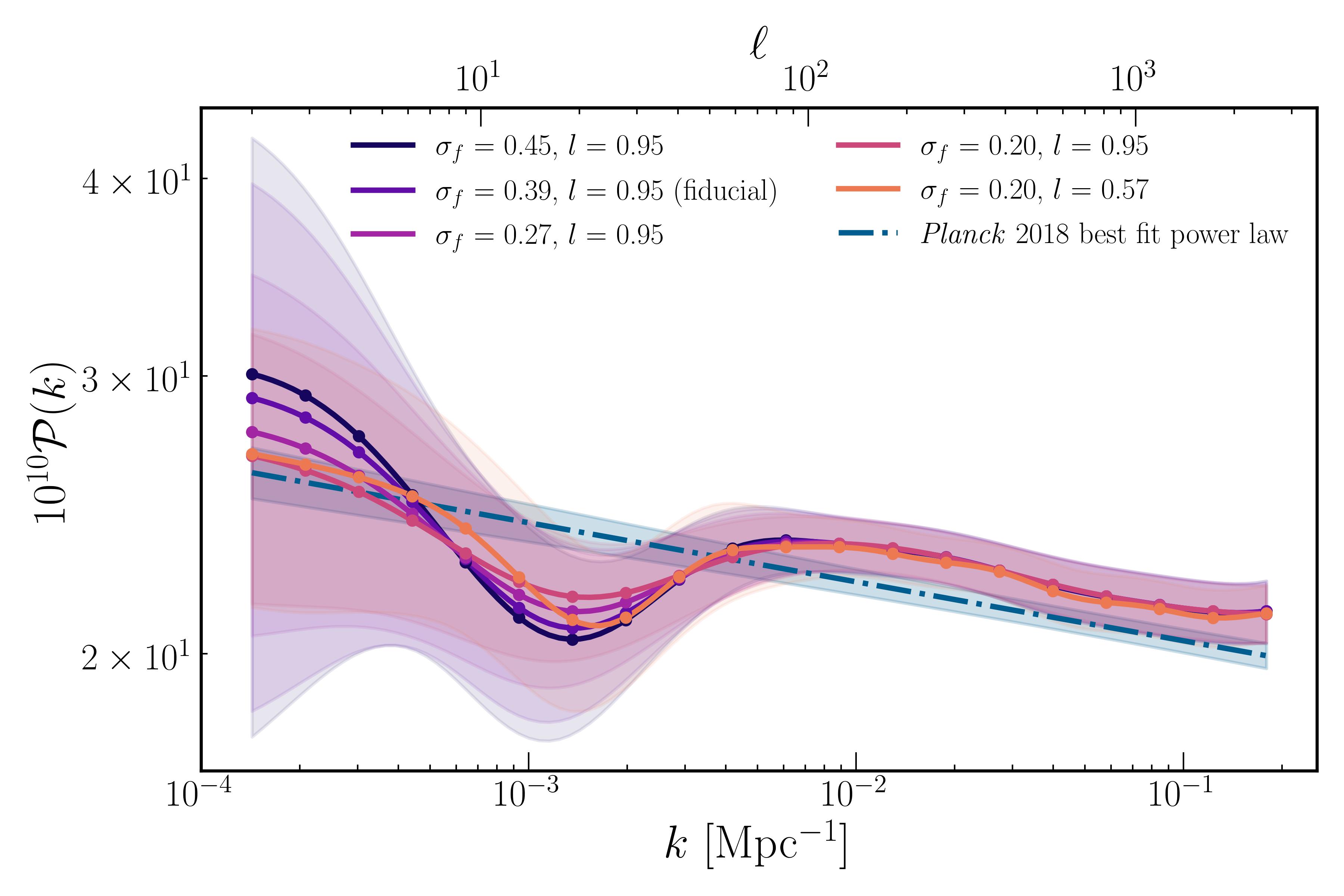}
\caption{\label{fig:hyperparams_change_recons}The mean reconstructed PPS for different GP hyperparameter settings. The shaded regions around the curves represent the 68\% confidence intervals.}
\end{figure*}
As discussed in Section \ref{subsec:p(k)}, the properties of GPs are governed by two hyperparameters, the length scale $l$ and the signal standard deviation $\sigma_f$, which we have fixed in our fiducial analysis. In order to ensure that our results are not driven by our particular choice of these hyperparameters, we repeat our analysis for a number of different hyperparameter choices. Specifically, we train several emulators with different values for $l$ and $\sigma_f$, while keeping all other settings consistent with those used in our fiducial analysis. The joint parameter constraints obtained for the four different cases considered are shown in Appendix~\ref{appendix:n(p(k))}, with the corresponding reconstructed PPSa in Fig.~\ref{fig:hyperparams_change_recons}. As can be seen, the mean reconstructed PPSa remain consistent for all cases considered. In particular, the nearly scale-invariant form of the reconstructed PPS at small scales is robust to any hyperparameter changes, as is the observed lack of power around $k \sim 1.57 \times 10^{-3}$ $\text{Mpc}^{-1}$. The only difference observed is that the uncertainties on the reconstructed PPSa on large scales decrease as we decrease $\sigma_f$. This is expected, since, as discussed above, the constraints at these scales are prior-driven and decreasing $\sigma_f$ effectively tightens the prior on the GP. The constraints on the small-scale PPSa are unaffected, as in this regime our contours are driven by the data rather than the prior. 

\begin{figure*}[tbp]
\centering % \begin{center}/\end{center} takes some additional vertical space
\includegraphics[width=0.82\textwidth]{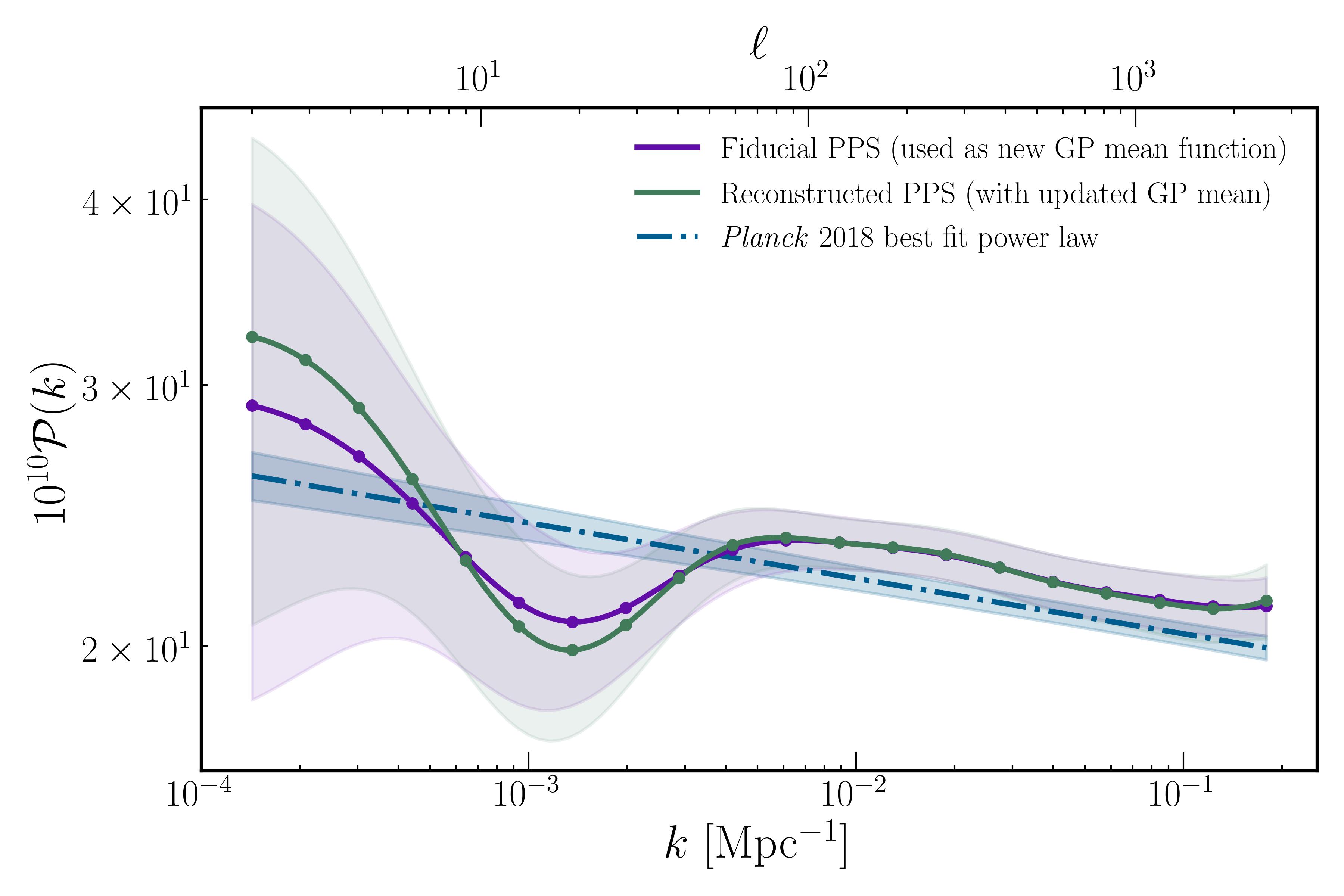}
\caption{\label{fig:mean_change}The fiducial reconstructed PPS is compared with the PPS reconstructed using the fiducial reconstruction as the mean function. The shaded region represents the $68\%$ confidence interval.}
\end{figure*}

To assess the robustness of our results against the choice of the GP mean function, we repeat our analysis using the reconstructed PPS from our fiducial results (which we obtain via the differentiable framework) as the new mean function. The resulting reconstructed PPS, shown in Fig.~\ref{fig:mean_change}, remains consistent with the fiducial results, with some deviations on the prior-dominated regime which are well within the statistical uncertainties. On small scales, the reconstruction is fully data driven and consistent with the near–scale-invariant behaviour observed in the fiducial results. This demonstrates that our findings are not biased by the choice of the GP mean function, but are instead strongly driven by the data across the considered scales. An extension of this analysis would involve adopting the approach introduced in Ref.~\cite{2022MNRAS.512.1967R}, where a free amplitude parameter is introduced to scale the GP mean function. This parametrization allows the mean itself to be constrained by the GP. We leave this more comprehensive investigation for future work.

Finally, the number of nodes with which we parametrize the GP represents an additional hyperparameter in our analysis. As we show in Appendix \ref{appendix:n(p(k))}, we find our results to not depend on this choice, and thus conclude that our main results are robust to changes in the GP hyperparameters and the mean function, with only minor differences in the derived uncertainties in the prior-dominated regime.

\section{\label{sec:conclusions}Conclusions}

In this work, we present a novel, fully-differentiable pipeline to reconstruct the primordial power spectrum (PPS) generated from inflation. We model the PPS using a Gaussian process (GP), which in our fiducial setup we parametrize through a set of $n=20$ GP nodes, and obtain theoretical predictions for cosmological observables by training a neural network emulator based on the Boltzmann code \textsc{camb} and using the \textsc{CosmoPower} framework. To ensure differentiability, we implement the likelihood entirely in \texttt{JAX}, in particular making use of \texttt{Cosmopower-JAX}. This approach allows us to employ gradient-based sampling algorithms, particularly the No-U-Turn-Sampler extension of Hamiltonian Monte Carlo (HMC), to efficiently explore the parameter space spanned by the cosmological parameters and the Gaussian process nodes.

We then apply this pipeline to constrain the PPS using CMB temperature anisotropy data from \textit{Planck} 2018, alongside a Gaussian prior on the optical depth to reionization based on \textit{Planck} 2015 low-$\ell$ polarisation data. On small scales, our reconstructed PPS is consistent with near scale-invariance as predicted from single-field slow-roll inflation. On large scales, our constraints remain relatively weak, mostly due to cosmic variance. Nonetheless, we recover the well-known dip in the CMB temperature power spectrum at around $\ell=22$ as a corresponding dip in the reconstructed PPS. Comparing to the $\Lambda$CDM results from \textit{Planck} Collaboration 2018, we find the PPS reconstructed in this work to exhibit a slightly higher amplitude on small scales. This can be attributed to our choice of a wider prior on the optical depth to reionization $\tau$, with our fiducial results based on the measurement from \textit{Planck} 2015, which was later revised downwards in \textit{Planck} 2018 due to the measurements from the High-Frequency instrument. Incorporating these latter measurements into our analysis brings our results into full agreement with the results from \textit{Planck} 2018, which is due to the degeneracy between the primordial power spectrum amplitude and $\tau$, such that lower optical depth leads to lower amplitude.

Finally, we test the robustness of our results by performing a suite of consistency tests. In particular, we compare our HMC-based results to results based on importance nested sampling, and we repeat our analysis using different GP hyperparameter choices. We find consistent results for all these analysis variants, thus demonstrating the robustness of our results.

The flexible and efficient pipeline developed in this work is well-suited for application to additional datasets. A natural extension of this work would be to incorporate CMB polarisation anisotropy data, thus mitigating the dependence of our results on the imposed $\tau$-prior. Improved measurements of the optical depth, such as those recently reported by the CLASS telescope~\cite{2025arXiv250111904L}, are expected to significantly enhance the precision of PPS reconstructions, particularly at low multipoles. Our analysis framework also allows for extensions beyond CMB data by incorporating large-scale structure observables such as cosmic shear or galaxy clustering. Therefore, this work lays the groundwork for multi-probe primordial power spectrum reconstruction, which will be crucial for obtaining tight and robust constraints on inflation using current as well as upcoming data. Furthermore, the methodology developed as part of this work can also be employed for future matter power spectrum reconstructions~\cite[e.g.][]{Preston:2024ggf, Broxterman:2024oay, Ye:2024rzp, simon2025kids1000detectiondeviationspurely}.

\appendix
\section{\label{appendix:n(p(k))}Validating changes in GP hyperparameters}

\subsection{Effect of changing the number of nodes of the GP}
\begin{figure*}[tbp]
  \centering
  \begin{subfigure}{0.495\textwidth}
    \centering
    \includegraphics[width=\linewidth]{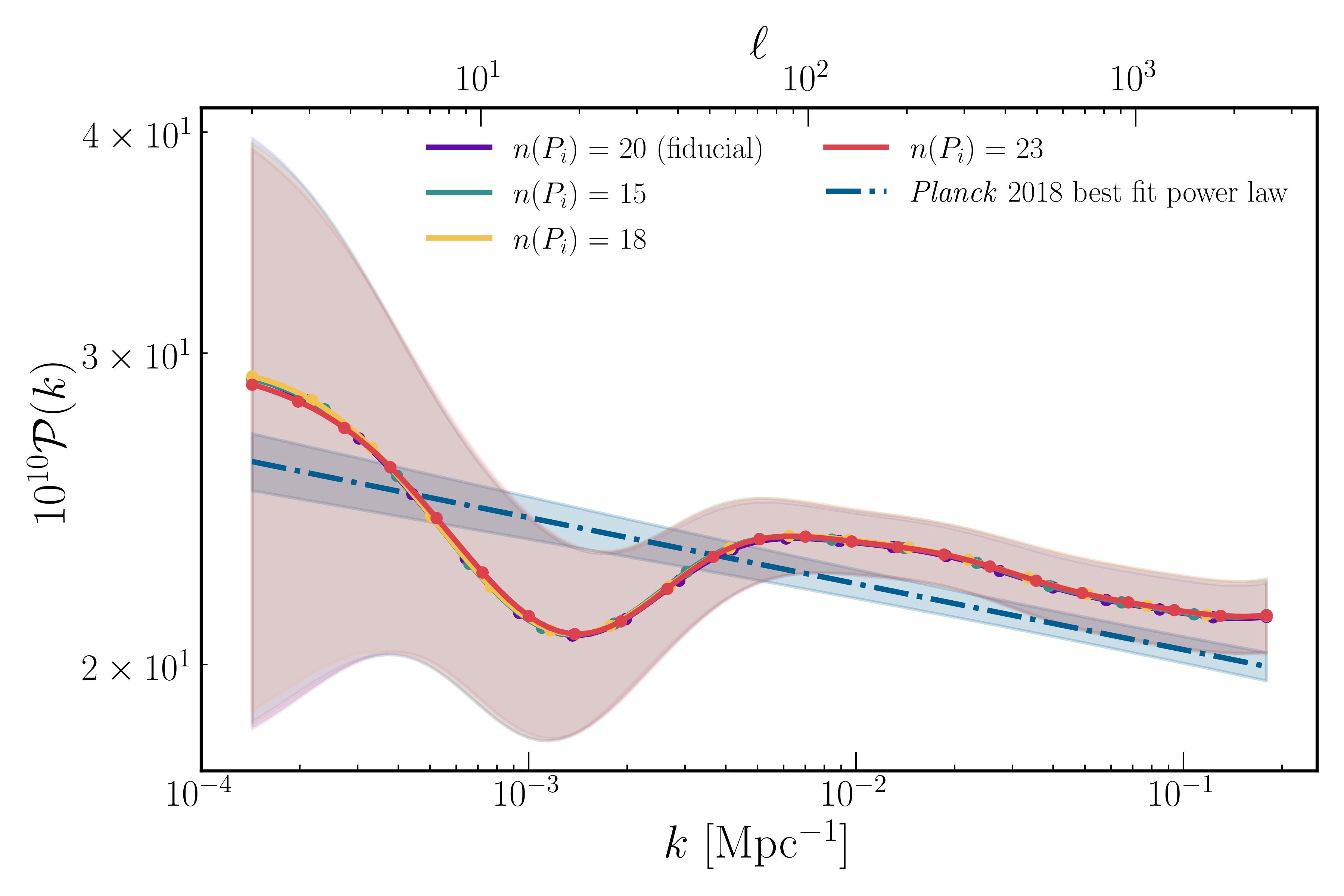}
    \caption{}
  \end{subfigure}
  \hfill
  \begin{subfigure}{0.495\textwidth}
    \centering
    \includegraphics[width=\linewidth]{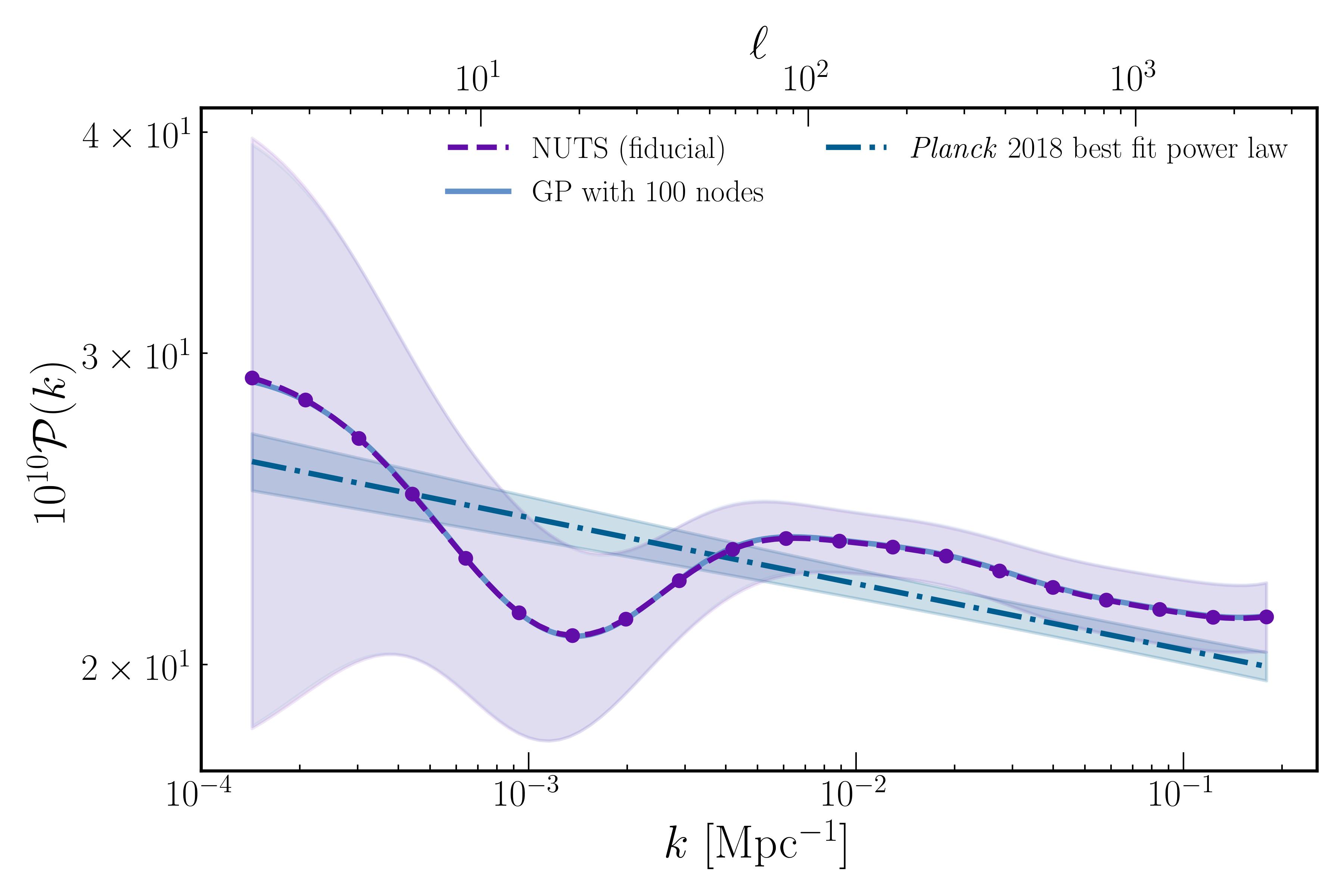}
    \caption{}
  \end{subfigure}
  \caption{Reconstructed PPS shown with solid lines, and corresponding $68\%$ confidence intervals indicated by shaded regions. (a) PPS reconstructions for different numbers of GP nodes across each $k$ mode. (b) Comparison of the PPS reconstructed using 100 GP nodes with the fiducial reconstruction.}
  \label{fig:diff_n(pK)}
\end{figure*}

We present the reconstructed PPSa obtained by varying the number of nodes in the GP framework (see Sec.~\ref{subsec:p(k)}) while keeping the GP hyperparameters fixed to the values used in our fiducial analysis. As shown in the left panel of Fig.~\ref{fig:diff_n(pK)}, changes in the number of nodes do not significantly affect the reconstruction—the PPS remains largely consistent across different configurations.

However, the choice of the number of GP nodes has important implications. With fixed hyperparameters, there exists a lower limit on the number of nodes required for a stable reconstruction. We find that reducing the number of nodes excessively, without adjusting the GP’s amplitude ($\sigma_f$) and smoothness ($l$), can lead to irregular, highly fluctuating spectra. This irregularity can impair the NN emulator’s ability to learn the mapping between input parameters and output features effectively.

Conversely, increasing the number of nodes, when combined with well-calibrated hyperparameters, raises the dimensionality of the problem. However, our fully differentiable framework is designed to efficiently manage this increased complexity while preserving robustness and scalability. To evaluate the consistency of our approach under more extreme conditions, we increased the number of nodes to 100 (see right panel of Fig.~\ref{fig:diff_n(pK)}), while keeping all other hyperparameters fixed at their fiducial values; this expansion raises the total number of free parameters to 105. This high number of parameters showcases the potential for gradient-based sampling and differentiable programming to scale the inference to the complexity required by next-generation cosmological surveys \citep{CPJ2023OJAp....6E..20P, 2024OJAp....7E..73P}.

With a higher node count, one might expect the reconstruction to resolve finer features that were previously smoothed out. However, our results show that the reconstructed PPS remains consistent with the fiducial analysis, suggesting that no additional features can be extracted within the constraints of our current GP model and hyperparameter settings. A potential avenue for future investigation is to explore models with correlation lengths smaller than the node separation, which may reveal additional structure.

\subsection{Effect of changing the GP hyperparameters}
\begin{figure*}[tbp]
\centering % \begin{center}/\end{center} takes some additional vertical space
\includegraphics[width=1.0\textwidth]{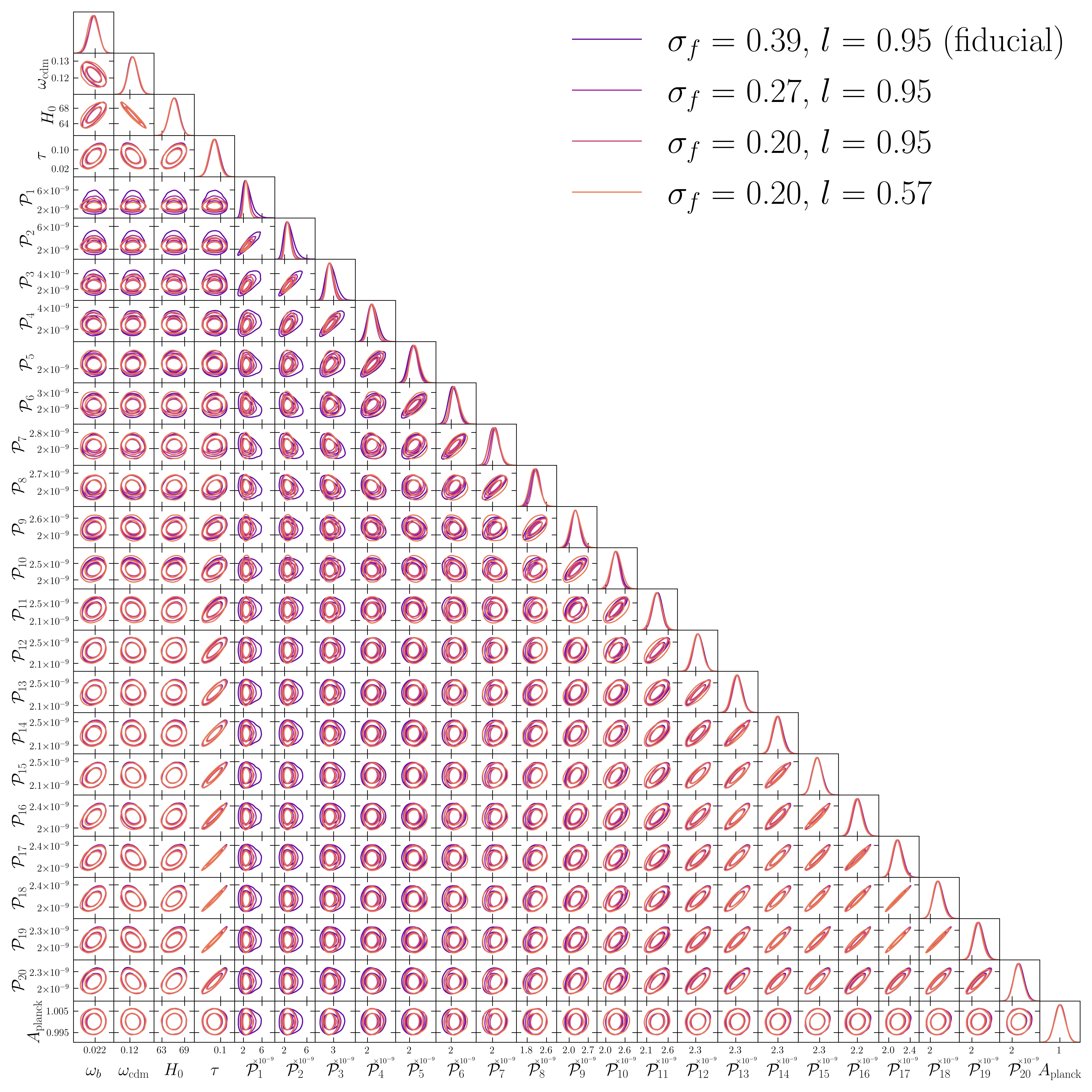}
\caption{\label{fig:hyperparams_change}Constraints on cosmological and nuisance parameters, as well as PPS values at each $k$ mode, obtained for different GP hyperparameter settings of $l$ and $\sigma_f$, as indicated in the legend. The contours represent the 68\% and 95\% confidence regions, accompanied by 1-D marginal distributions for each parameter.}
\end{figure*}

In Fig.~\ref{fig:hyperparams_change}, we present the joint constraints on the cosmological parameters, the PPS values at each $k$ mode, and the nuisance parameter. The constraints on the cosmological parameters remain consistent across all hyperparameter settings. As discussed in the main text and compared with the reconstruction in Fig.~\ref{fig:hyperparams_change_recons}, we find that the contours for small-scale $\mathcal{P}_i$ remain stable across different hyperparameter choices, aligning with the reconstructed PPS. On large scales, the variations are also consistent with the reconstruction, with the contours shrinking as $\sigma_f$ is lowered.

\section{\label{appendix:nn_train}Neural network training and performance}

We maintain the CP architecture consistent with the original implementation, with only minor modifications. One notable change is implementing a feature to iteratively save the training and validation losses for each learning rate. We also adjust the training loop to iterate over multiple learning rates, starting with larger updates in the initial stages to facilitate faster convergence and gradually decreasing the learning rate in later epochs to fine-tune the model for improved accuracy. We also increase the \texttt{batch\_size} to 1024 and the \texttt{max\_epochs} to 1500, providing the model with sufficient training time if needed. An early stopping mechanism is incorporated in the original framework to terminate training when further improvements are unlikely, ensuring computational efficiency. We trained our NN emulators on an NVIDIA A40 GPU, which significantly improved computational efficiency compared to CPU-based training. Although we experimented with training on a CPU, the GPU's parallel processing capabilities dramatically reduced training time and enhanced overall performance.

\begin{figure*}[tbp]
\centering % \begin{center}/\end{center} takes some additional vertical space
\includegraphics[width=0.80\textwidth]{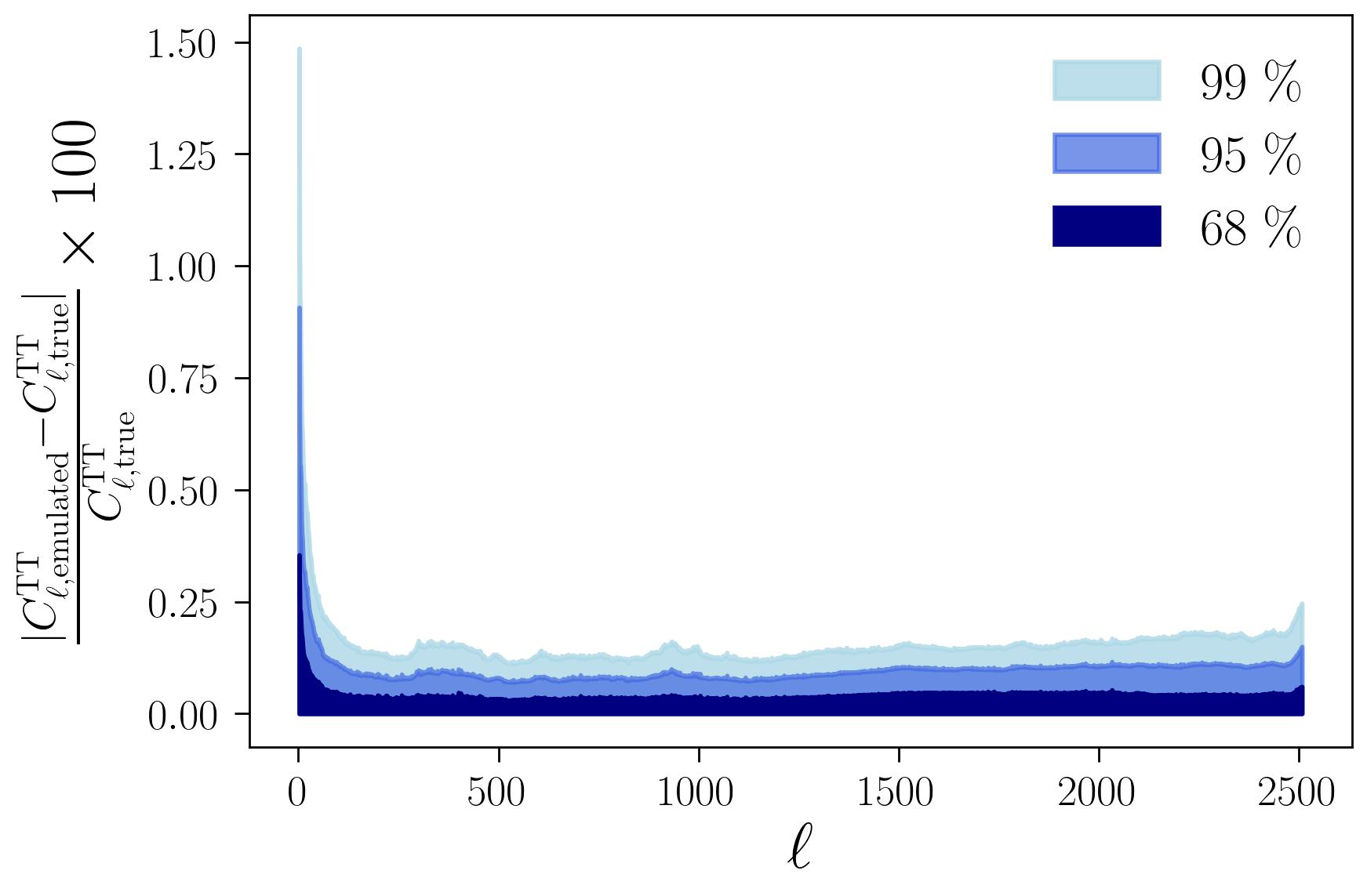}
\caption{\label{fig:emulation_error}Emulation accuracy of the CMB power spectra emulator, calculated in terms of relative error. The different colours indicate regions corresponding to the 68th, 95th, and 99th
percentiles of the test set as evaluated by the emulator.}
\end{figure*}

In Figure \ref{fig:emulation_error}, we illustrate the performance of our trained emulator in terms of relative error. Across the entire multipole range, the relative error in emulation accuracy remains below \(1.5\%\), demonstrating the emulator’s capability to accurately reproduce any spectra within the training prior range specified in Table \ref{tab:prior_tab}. 
When using \textsc{camb} to generate the corresponding angular power spectrum, we set the parameters \texttt{lmax} and \texttt{lens\_potential\_accuracy} to 2509 and 1, respectively. The latter ensures \textit{Planck}-level accuracy for the lensing potential. As we are using \textit{Planck} 2018 data, we do not need to adjust additional accuracy parameters. We have checked the effect of using higher accuracy settings for our analysis and we did not find any significant improvements.

\section{\label{appendix:sampling}Sampling}

\subsection{Effect of imposed \texorpdfstring{$\tau$}{tau} prior}

\begin{figure*}[tbp]
\centering % \begin{center}/\end{center} takes some additional vertical space
\includegraphics[width=1.0\textwidth]{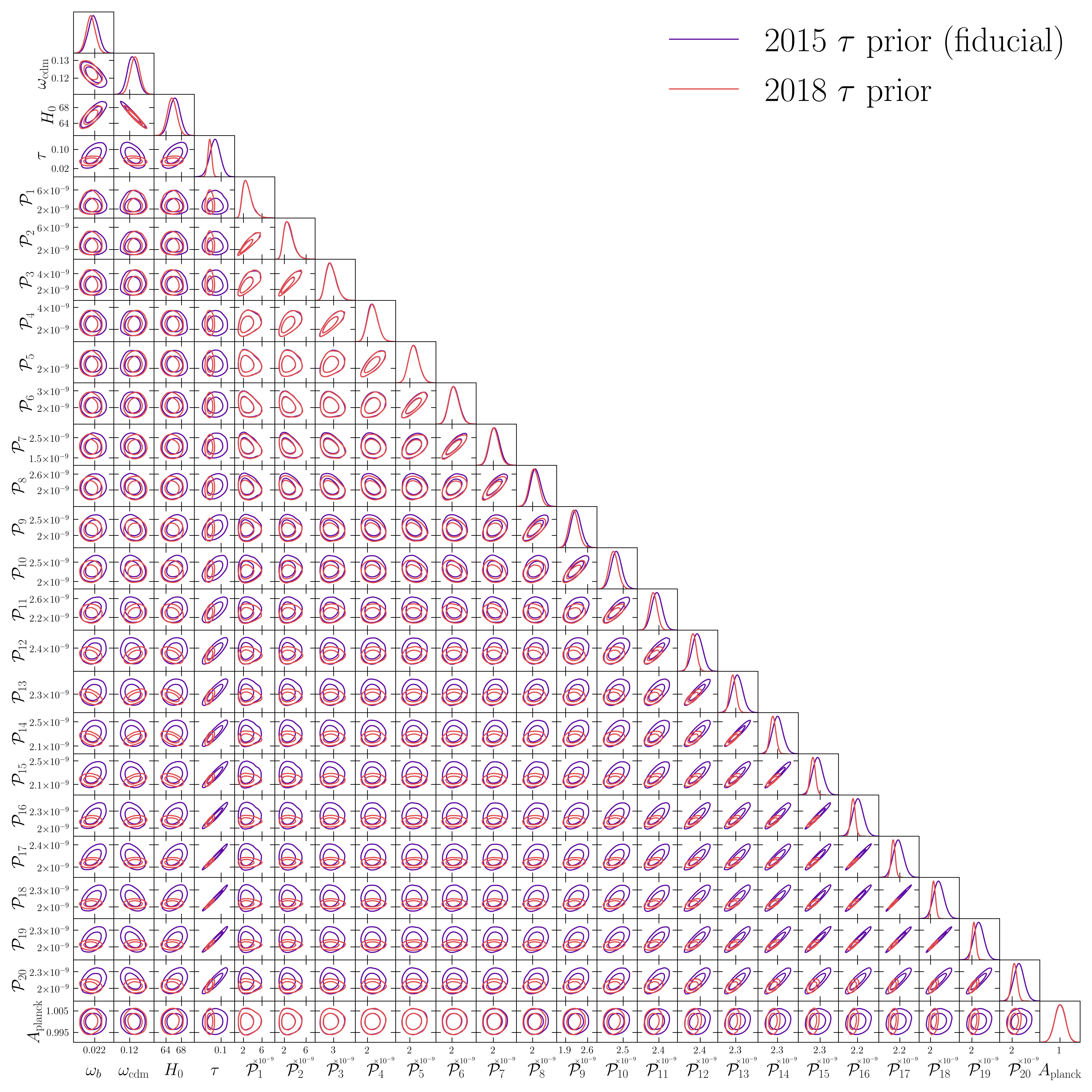}
\caption{\label{fig:diff_tau_full}The joint constraints on 20 nodes of the GP used to generate the PPS values at each $k$ mode, the cosmological parameters and the nuisance parameter using different $\tau$ priors. The contours represent the 68\% and 95\% confidence regions, accompanied by 1-D marginal distributions for each parameter.}
\end{figure*}
\begin{figure*}[tbp]
\centering % \begin{center}/\end{center} takes some additional vertical space
\includegraphics[width=0.63\textwidth]{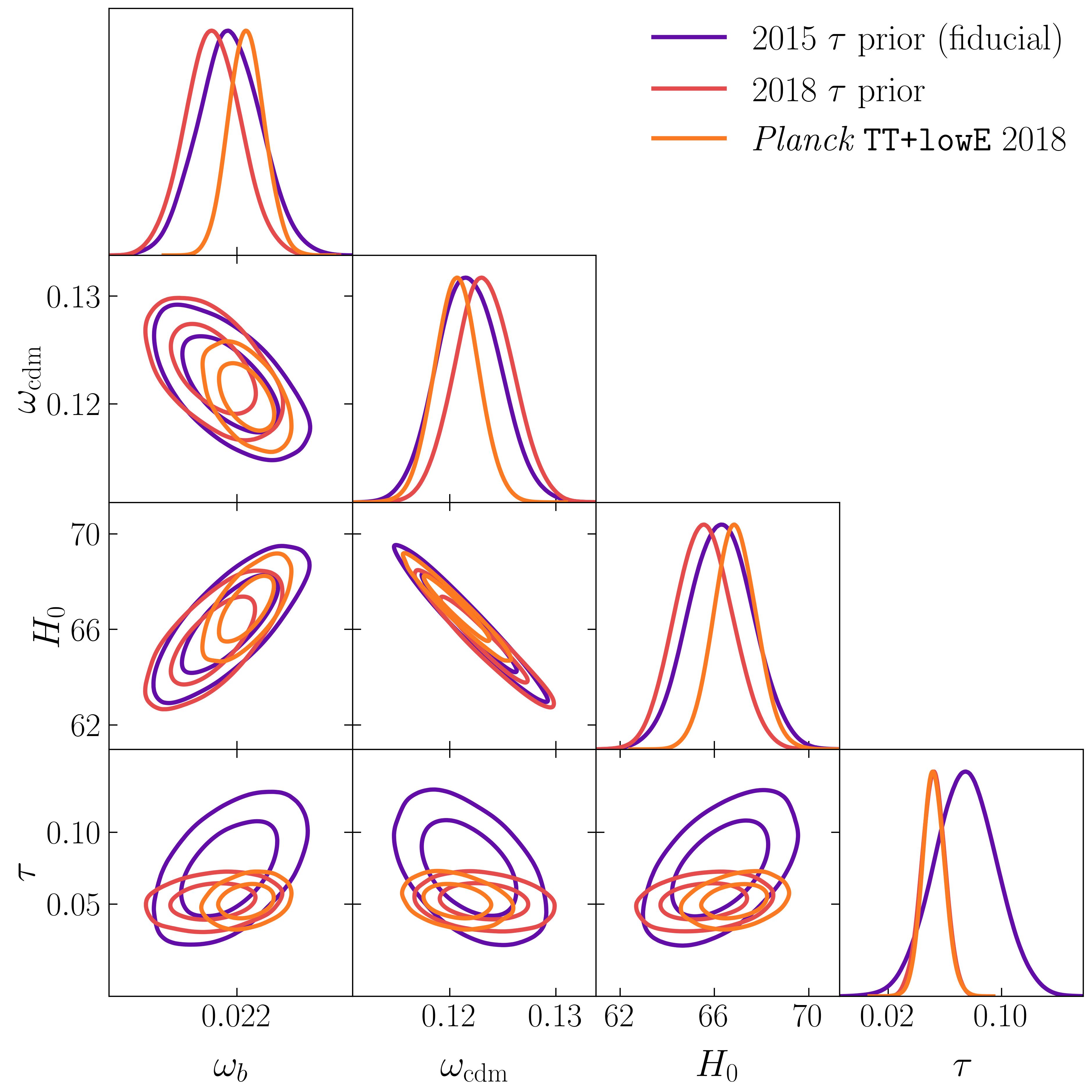}
\caption{\label{fig:planck_tau}Comparison of cosmological parameter constraints obtained using the updated 2018 $\tau$ prior with the \textit{Planck} 2018 \texttt{TT+lowE} results. The PPS and nuisance parameter have been marginalized over. The contours represent the $68\%$ and $95\%$ confidence intervals, with 1-D marginals for each parameter.} 
\end{figure*}
In Fig. \ref{fig:diff_tau_full}, we present the results of our joint constraints on cosmological parameters, PPS values at specific $k$ modes, and the nuisance parameter for the updated $\tau$ prior, based on \textit{Planck} 2018 low-$\ell$ polarisation data, compared to the fiducial results. Except for $\tau$, all cosmological parameters remain consistent within 1$\sigma$. As discussed in the main text, the observed shift in $\tau$ arises from the updated prior informed by \textit{Planck} 2018~\cite{2019arXiv190712875P}. Comparing this figure with Fig. \ref{fig:diff_tau}, we find that constraints on $\mathcal{P}_i$ at large scales remain unchanged, as seen in the overlapping contours in Fig.~\ref{fig:diff_tau_full}. This is expected since large-scale constraints are cosmic variance-limited, and \textit{Planck} 2018 did not significantly improve them over \textit{Planck} 2015. At smaller scales, however, the contours for $\mathcal{P}_i$ become tighter, reflecting a closer match to the \textit{Planck} 2018 best-fit power-law model. This trend is also apparent in Fig. \ref{fig:diff_tau} in the main text. The impact of the updated prior on $\tau$ is also illustrated in Fig.~\ref{fig:planck_tau}, where we compare our inferred constraints on the cosmological parameters to the \textit{Planck} 2018 \texttt{TT+lowE} results. With the updated prior, we infer $\tau = 0.0520 \pm 0.0083$, which is in good agreement with both the \textit{Planck} 2018 results and the latest large-scale CMB E-mode polarization measurements from the ground-based \textit{CLASS} telescope~\cite{2025arXiv250111904L}. We emphasize that the differences in the inferred values of $\omega_b$, $\omega_{\mathrm{cdm}}$, and $H_0$ between our reconstruction (with the updated $\tau$ prior) and the \textit{Planck} 2018 \texttt{TT+lowE} results are consistent with those reported in the Planck analysis of PPS reconstructions (see Figure 18 of Ref.~\cite{2020A&A...641A..10P}).

\subsection{Testing robustness of sampling techniques}
To ensure the robustness of our \textsc{INS}-based results, we conduct a series of tests by varying key hyperparameters in the \textsc{Nautilus} implementation. Specifically, we explore different values for the number of live points, $n_{\mathrm{live}}$, and the effective sample size, $N_{\mathrm{eff}}$, which control the precision and efficiency of the sampling process. Across all hyperparameter settings considered, we obtain consistent results, demonstrating that constraints are robust and not artifacts of specific hyperparameter choices.

We also test the robustness of our \textsc{NUTS}-based results by varying the $\texttt{target\_accept\_prob}$ hyperparameter, which controls the target acceptance probability for step size adaptation. Increasing this parameter leads to a smaller step size, making the sampling process slower but more robust by reducing the likelihood of divergences.

The default value for $\texttt{target\_accept\_prob}$ is 0.8. We incrementally increase this value in our tests and observed no significant changes in the results, indicating that our analysis is robust to this hyperparameter choice and unaffected by divergences. We specifically focused on this hyperparameter because divergences in HMC can signal poor exploration of the posterior distribution. The consistency of our results across different $\texttt{target\_accept\_prob}$ values reassures us that our sampling accurately captures the posterior without being influenced by numerical artifacts or parameter tuning.

\subsection{Details of sampler specific hyperparameters for fiducial results}
The joint constraints in Fig.~\ref{fig:nuts_ns1} result from two different sampling approaches: the primary sampling algorithm, \textsc{NUTS}, within the differentiable pipeline, and with the non-differentiable framework using \textsc{INS}.

For the \textsc{NUTS} sampler, we perform the analysis on a high-performance computing system, utilizing four NVIDIA A40 GPUs. The MCMC setup with a \textsc{NUTS} kernel is configured with 1000 warm-up iterations, 4000 samples, and four chains, initializing the parameters to the median of the prior distribution. We did not specify any value for the $\texttt{inverse\_mass\_matrix}$ and thus it was initialised to the identity matrix by default. This configuration allows us to complete the sampling process in approximately 2 hours.

In comparison, we conduct the same analysis using the non-differentiable \textsc{INS} framework. Here, we configure the sampler with 3000 live points ($n_{\mathrm{live}}$) and a pool of 48 CPU cores. The sampler runs with 8000 effective samples ($N_{\mathrm{eff}}$), discarding exploration to improve convergence. This setup takes significantly longer, approximately 1 day and 17 hours.

By comparing the results, we highlight the efficiency of the differentiable \textsc{NUTS} pipeline, which benefits from GPU acceleration and gradient-based optimization methods, in contrast to the more computationally intensive \textsc{INS} approach that relies solely on CPU cores.

\acknowledgments
ASM thanks Princeton University and the Flatiron Institute for their hospitality during his visit, where this project was initiated. DP was supported by the SNF Sinergia grant CRSII5-193826 “AstroSignals: A New Window on the Universe, with the New Generation of Large Radio-Astronomy Facilities.'' We also thank the anonymous referee for their insightful comments and feedback.


%\begin{thebibliography}{99}
% \bibliographystyle{JHEP}
% \bibliography{references}

\providecommand{\href}[2]{#2}\begingroup\raggedright\endgroup

% Please avoid comments such as "For a review'', "For some examples",
% "and references therein" or move them in the text. In general,
% please leave only references in the bibliography and move all
% accessory text in footnotes.

% Also, please have only one work for each \bibitem.

\end{document}